\documentclass[prl,aps,twocolumn,superscriptaddress]{revtex4}
\usepackage{epsfig}
\usepackage{amsmath}
\input{epsf}
\usepackage{psfrag}
\usepackage{xcolor}
\usepackage{slashed}

\usepackage{ulem} 
\usepackage{soul}
\usepackage{CJKutf8}
\usepackage{hyperref}

\DeclareRobustCommand\old{\bgroup\markoverwith{\color{magenta}{\rule[0.4ex]{2pt}{0.8pt}}}\ULon}

\newcommand{\nn}{\nonumber}
\newcommand{\Tau}{{\cal T}}
\newcommand{\as}{{\alpha_s}}
\newcommand{\cusp}{{\rm cusp}}
\def\d{{\rm d}}
\begin{document}


\title{Nucleon Tomography with 0-jettiness }

\author{Shen Fang}
\email{sfang23@m.fudan.edu.cn}
\affiliation{Department of Physics, Center for Field Theory and Particle Physics, Key Laboratory of Nuclear Physics and Ion-beam Application (MOE), Fudan University, Shanghai 200433, China \vspace{0.1cm}}

\author{Shuo Lin}
\email{shuolin@sdu.edu.cn}
\affiliation{Key Laboratory of
Particle Physics and Particle Irradiation (MOE), Institute of
Frontier and Interdisciplinary Science, Shandong University,
(Qingdao), Shandong 266237, China \vspace{0.1cm}}

\author{Ding Yu Shao}
\email{dingyu.shao@cern.ch}
\affiliation{Department of Physics, Center for Field Theory and Particle Physics, Key Laboratory of Nuclear Physics and Ion-beam Application (MOE), Fudan University, Shanghai 200433, China \vspace{0.1cm}}
\affiliation{Shanghai Research Center for Theoretical Nuclear Physics, NSFC and Fudan University, Shanghai 200438, China \vspace{0.1cm}}

\author{Jian Zhou}
\email{jzhou@sdu.edu.cn}
\affiliation{Key Laboratory of
Particle Physics and Particle Irradiation (MOE), Institute of
Frontier and Interdisciplinary Science, Shandong University,
(Qingdao), Shandong 266237, China \vspace{0.1cm}}
\affiliation{Southern Center for Nuclear-Science Theory (SCNT), Institute of Modern Physics, Chinese Academy of Sciences, Huizhou, Guangdong
516000, China\vspace{0.1cm}}

\begin{abstract}

We propose a novel strategy to systematically isolate the nucleon's intrinsic non-perturbative three-dimensional structure by employing 0-jettiness to suppress initial-state radiation in transverse-momentum-dependent observables. Applying this method to transverse single spin asymmetries (SSAs) in $W^\pm$ and $Z^0$ boson production at RHIC, we demonstrate a substantial enhancement of the asymmetry signal (e.g., by $115\%$ for $W^-$ SSA at $q_\perp=5$ GeV). We show that this enhancement yields a substantial net gain in experimental sensitivity—even after accounting for the statistical cost of the veto—facilitating a more definitive test of the predicted Sivers function sign change. We further explore its applicability to spin-dependent measurements at the Electron-Ion Collider. Our analysis is formulated within a joint resummation framework that systematically resums large logarithms associated with both the veto scale and the gauge boson's transverse momentum.
 
\end{abstract}

\maketitle

\date{\today}

{\it Introduction} --- 
Understanding the three-dimensional (3D) structure of the nucleon is of fundamental importance to hadron physics, particularly with the upcoming Electron-Ion Colliders (EICs) in the United States~\cite{Accardi:2012qut, AbdulKhalek:2021gbh} and China~\cite{Anderle:2021wcy}.  One of the central objectives of these future facilities is to map out the 3D motion of partons inside the nucleon through $ep$ and $eA$ collisions, as characterized by transverse-momentum-dependent (TMD) parton distributions~\cite{Collins:1981uk, Collins:1981uw, Ji:2004wu, Collins:2011zzd, Boussarie:2023izj}. TMD distributions not only provide a direct 3D imaging of the nucleon~\cite{Mulders:1995dh, Mulders:2000sh}, but also offer insights into the nucleon’s spin structure~\cite{Sivers:1989cc, Collins:1992kk, Mulders:1995dh, Mulders:2000sh, Liang:1997rt, Ji:2002xn} and shed light on fundamental aspects of QCD factorization theorem~\cite{Brodsky:2002cx, Brodsky:2002rv, Collins:2002kn, Collins:2004nx, Ji:2002aa}. They also encode crucial information on the color glass condensate---a unique form of nuclear matter at small $x$~\cite{McLerran:1993ni, McLerran:1993ka, Dominguez:2011wm,  Metz:2011wb, Akcakaya:2012si, Zhou:2013gsa, Kotko:2015ura, Boer:2015pni,  Balitsky:2016dgz, Altinoluk:2019wyu}.

A significant challenge in probing the intrinsic transverse motion of partons is the contamination from initial-state radiation (ISR), whose recoil effects introduce perturbative contributions that obscure and complicate the disentanglement of the non-perturbative structure of TMDs~\cite{Boussarie:2023izj}. To address this, we propose in this Letter a novel strategy: employing 0-jettiness as a veto to suppress central ISR. This approach enhances experimental sensitivity to the inherent non-perturbative TMD dynamics, opening new avenues to separate these from perturbative contamination, a persistent difficulty that has limited precision extractions at collider energies. Notably, while semi-inclusive deep inelastic scattering (SIDIS) processes allow for experimental tuning of the hard scale $Q$, electroweak Drell-Yan processes have $Q$ fixed by the gauge boson mass. Here, 0-jettiness emerges as an effective tunable energy scale, offering novel flexibility in controlling TMD evolution effects.

\begin{figure}
\centering
  \includegraphics[scale=0.25]{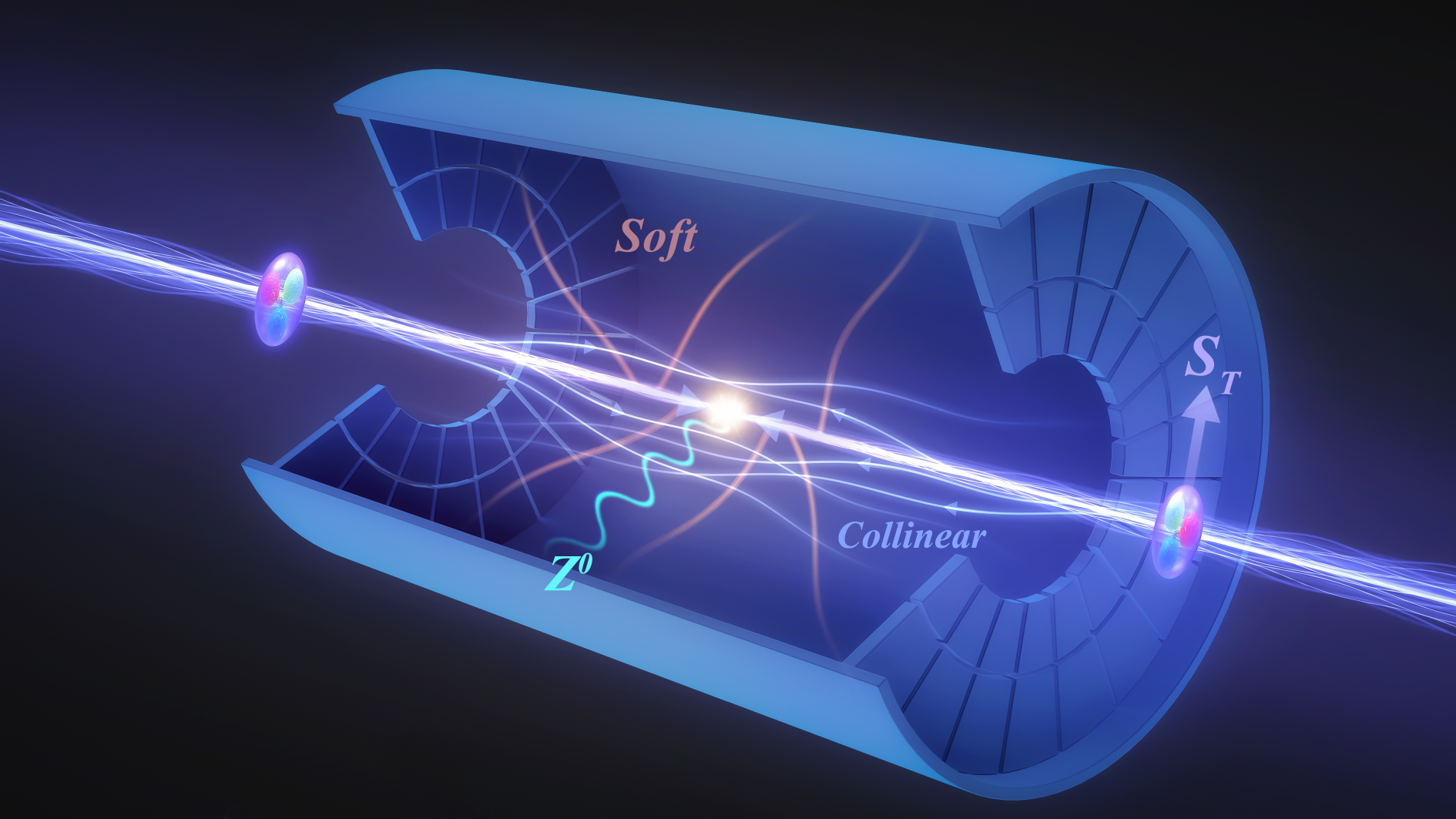}
\caption{Schematic illustration of the final state configuration for $Z^0$ production in polarized $pp$ collisions with 0-jettiness. The faint red lines represent the soft radiation vetoed by the cut, and the bright straight lines represent the collinear hard radiation that drives the scale evolution of PDFs and gives rise to SSAs via the Qiu-Sterman mechanism.  }
\label{fig:Z0}
\end{figure}

The 0-jettiness observable, originally introduced as a global event shape to veto additional jets in an inclusive manner~\cite{Stewart:2010tn}, has become a powerful tool in collider phenomenology~\cite{Gaunt:2015pea, Boughezal:2015aha, Boughezal:2015dva, Boughezal:2016wmq}. This observable enhances signal sensitivity by effectively suppressing additional radiation. Recent advances in resummation techniques have enabled precise and systematically improvable predictions for 0-jettiness cross sections~\cite{Stewart:2010pd, Kang:2013lga, Lustermans:2019plv, Alioli:2021ggd, Alioli:2023rxx, Knobbe:2023ehi, Cao:2024ota}, thereby playing an essential role in interpreting LHC data.

Despite these advances, the application of the 0-jettiness veto to polarization-dependent TMD observables remains unexplored. Here, we present the first-ever extension of joint resummation~\cite{Jain:2011iu, Procura:2014cba, Monni:2019yyr, Lustermans:2019plv, Makris:2020ltr} to polarization-dependent observables in Drell-Yan-like processes shown in Fig.~\ref{fig:Z0}, combining veto-induced logarithms and $k_t$-logarithms. More specifically, we investigate the impact of the 0-jettiness veto on transverse single spin asymmetries (SSAs) in $W^\pm$ and $Z^0$ boson production at RHIC~\cite{Kang:2009bp, Metz:2010xs}, which provides a critical test of the predicted sign change of the Sivers function~\cite{Collins:2002kn}. Existing SSA measurements are strongly diluted by TMD evolution effects at high energies~\cite{Kang:2011mr, Echevarria:2020hpy}. We demonstrate that imposing a central 0-jettiness veto significantly enhances the SSA signal generated by the Qiu-Sterman mechanism~\cite{Qiu:1991pp, Ji:2006ub}. Crucially, we show that this enhancement provides a substantial net gain in experimental sensitivity, even after accounting for the statistical cost of the veto. Moreover, while our results focus on transverse spin asymmetries in Drell-Yan processes, the formalism is broadly applicable. This approach opens a new strategy to enhance sensitivity to intrinsic parton motion across a wide range of TMD observables, with relevance for RHIC, LHC, the future EIC, and beyond.


{\it TMDs with 0-jettiness} --- As TMD evolution is predominantly driven by soft gluon radiation, observables that suppress radiation in the central rapidity region are particularly effective for enhancing sensitivity to non-perturbative effects. Among the various event shape observables, beam thrust, also referred to as 0-jettiness~\cite{Stewart:2009yx, Stewart:2010tn, Kang:2012zr, Jouttenus:2013hs,  Kang:2013wca, Gaunt:2014xga}, stands out as a powerful tool for this purpose. For electroweak Drell-Yan processes, the 0-jettiness variable is defined as
\begin{align} \label{eq:beam_thrust}
\tau&\equiv  \frac{2}{Q^2}  \sum_i \min \{p_a \cdot l_i,p_b \cdot l_i\}=\sum_i \frac{|\vec l_{\perp,i}|}{Q}e^{-|y_i-y|}\,,
\end{align}
where $Q$ and $y$ are the mass and rapidity of the produced gauge boson, respectively. The sum runs over all particles $i$ (excluding the gauge boson) with momentum $l_i$, transverse momentum $\vec l_{\perp,i}$ and rapidity $y_i$ in the final state. The momenta of the incoming partons are defined as $p_a^\mu=Q e^{+y} n_a^\mu/2$ and $p_b^\mu=Q e^{-y} n_b^\mu/2$, where $n_{a,b}^\mu=(1,0,0,\pm1)$ are light-like reference vectors. This observable is designed so that energetic particles collimated along the beam directions contribute minimally to $\tau$, while central emissions contribute significantly. Imposing a veto $\tau < \tau_0$ therefore strongly suppresses central gluon radiation and effectively constrains ISR effects.

We examine the transverse momentum spectrum of the gauge boson across a wider range of scales, where the perturbative expansion is spoiled by two distinct types of large logarithms: $\ln (Q^2/q_\perp^2)$ and $\ln (1/\tau_0)$, thereby going beyond the standard TMD resummation framework, which accounts only for the former.
While previous studies have explored the joint resummation of TMDs with 0-jettiness or jet veto~\cite{Jain:2011iu, Procura:2014cba, Monni:2019yyr, Lustermans:2019plv, Makris:2020ltr}, we present a novel derivation that is both transparent and heuristic, aiming to enhance accessibility to a broader audience. Our formulation unifies the various soft-collinear effective theory~(SCET)~\cite{Bauer:2000yr, Bauer:2001ct, Bauer:2001yt, Bauer:2002nz, Beneke:2002ph} regimes (SCET$_{\text{I}}$, SCET$_{\text{II}}$, and SCET$_{+}$) and is applicable in a more comprehensive kinematic region. 

{\it Sudakov form factor in TMD resummation} ---
To establish the joint resummation framework, we first recall the standard TMD factorization formula for the unpolarized Drell-Yan cross section \cite{Collins:1981uk, Collins:1981uw, Ji:2004wu, Collins:2011zzd},
\begin{align}
\label{eq:tmd_factorization_general}
&\frac{\d \sigma_{UU}}{ \d y \, \d^2 \vec{q}_\perp} = \sigma_0 \sum\limits_{q,q'} |V_{qq'}|^2 \int\frac{ \, \d^2 \vec b_\perp}{(2\pi)^2} e^{i \vec q_\perp \cdot \vec b_\perp} H_{qq'}(Q,\mu) \notag \\
&\qquad \times f_{q/p}(x_q,b,\mu, \zeta_q) f_{q'/p}(x_{q'},b,\mu, \zeta_{q'})\,,
\end{align}
where $b \equiv |\vec{b}_\perp|$, $\sigma_0$ is the Born cross section, $|V_{qq'}|^2$ is the CKM matrix element, $H$ is the hard factor, and $f_{i/p}(x_i, b, \mu, \zeta_i)$ is the unpolarized TMD PDF depending on scales $\mu$ and $\zeta$. Parton fractions are $x_q = Qe^y/\sqrt{s}$ and $x_{q'} = Qe^{-y}/\sqrt{s}$.

At NLL accuracy in the CSS framework \cite{Collins:1984kg, Collins:2011zzd}, for $q_\perp \gg \Lambda_{\text{QCD}}$, the TMD PDF is related to the collinear PDF $f_{q/p}(x, \mu_b)$ at $\mu_b = 2e^{-\gamma_E}/b$:
\begin{align}
\label{eq:tmd_to_coll}
f_{q/p}(x, b, \mu, \zeta) = e^{-S_{\rm P}^{\rm TMD}(b, \mu, \zeta)/2} f_{q/p}(x, \mu_b)\,.
\end{align}
Substituting into Eq.~\eqref{eq:tmd_factorization_general} with $\mu=Q$ and $\zeta_q=\zeta_{q'}=Q^2$, and using azimuthal integration, we obtain
\begin{align}
\label{eq:css_nll_cross_section}
&& \!\!\!\!\!\!\!\!\!\!\!\!\!\!\!\!
\frac{\d \sigma_{UU}}{ \d y \, \d^2 \vec{q}_\perp}= \sigma_0 \sum\limits_{q,q'} |V_{qq'}|^2 \int_0^\infty\frac{b \, \d b}{2\pi} J_0 (b\,q_\perp) \nonumber \\ &&
\times f_{q}(x_q,\mu_b)f_{q'}(x_{q'},\mu_b) e^{-S_{\rm P}^{\rm TMD}(b)}\,,
\end{align}
where $J_0$ is the Bessel function, and the perturbative Sudakov factor $S_{\rm P}^{\rm TMD}(b) \equiv S_{\rm P}^{\rm TMD}(b, Q, Q^2)$ is given by
\begin{align}
\label{eq:Sudakov_TMD}
S_{\text P}^{\rm TMD}(b)\!=\!\frac{C_F}{\pi} \int_{\mu_b^2}^{Q^2} 
\frac{\d \mu^2}{\mu^2} \left(
\ln \frac{ Q^2}{\mu^2}\!-\! \frac{3}{2}\right ) \alpha_s(\mu)\,.
\end{align}

{\it Sudakov form factor in joint resummation} ---
Imposing a veto $\tau < \tau_0$ restricts the phase space for ISR, introducing logarithms of $\tau_0$ and modifying the standard TMD Sudakov factor in Eq.~\eqref{eq:Sudakov_TMD}, particularly below the scale $t \equiv \tau_0 Q^2$. The resulting NLL perturbative Sudakov factor for the joint resummation, incorporating both TMD and 0-jettiness logarithms, is given by
\begin{eqnarray} \label{eq:Sud_joint}
&&\!\!\!\!\!\!\!\!\!\!\!\!
S_{\text P}(b)\!=\!\frac{C_F}{\pi}\! \left [\int_{\mu_b^2}^{t} \!\!
 \frac{\d \mu^2}{\mu^2}  
 \left ( 2\ln \frac{t}{\mu^2} -\frac{3}{2}\right ) -\int_{\mu_b^2}^{\tau_0t} \!\!
 \frac{\d \mu^2}{\mu^2} 
 \ln \frac{\tau_0 t}{\mu^2}  \nonumber \right .\  \\  &&
 \!\!\left .\     
 \!\!+\! \int_{t}^{Q^2} \!\!
 \frac{\d \mu^2}{\mu^2} \left(
 \ln \frac{ Q^2}{\mu^2}\!-\! \frac{3}{2}\right ) +\!\int^{t}_{\tau_0t} 
 \!\! \frac{\d \mu^2}{\mu^2} 
 \ln \frac{\mu^2}{ \tau_0t}   \right ]\!\alpha_s(\mu)\,. 
\end{eqnarray}
Here, the third integral, covering the range $[t, Q^2]$, corresponds to the standard TMD evolution above the scale $t$. The remaining terms encapsulate the modifications below the scale $t$ due to the phase space restrictions imposed by the 0-jettiness veto. This result is derived independently within both perturbative QCD phase space analysis and the SCET framework. Details are provided in Supplemental Material \cite{supplemental}, which includes Refs.~\cite{Collins:2011zzd,Procura:2014cba, Lustermans:2019plv,Stewart:2009yx}.

Our calculation employs NLL accuracy for the Sudakov resummation. While the NNLL joint resummation framework is available for the unpolarized cross section \cite{Lustermans:2019plv}, the corresponding one-loop corrections for polarized TMDs under a 0-jettiness veto remain unknown. Extending the formalism to NNLL accuracy is therefore beyond the scope of this Letter.


{\it Impact of 0-jettiness on spin asymmetries} --- We now examine the phenomenological implications of this approach for probing the spin structure of the nucleon. As an example, we focus on the SSAs for $W^\pm/Z^0$ production in polarized $pp$ collisions.  

As is well known, the scale evolution of polarization-dependent TMDs is governed by the standard Collins-Soper equation~\cite{Ji:2004xq}, since the light-cone divergence exhibits the same structure for both unpolarized and polarization-dependent cases. Accordingly, the analysis presented in the previous section extends directly to the polarization-dependent case, yielding the same perturbative Sudakov factor for the Sivers distribution. In the small $b$ limit (or equivalently, large $q_\perp$ region), the Sivers function can be further matched onto~\cite{Ji:2006ub} a collinear twist-3 quark-gluon correlation function, known as the Qiu-Sterman function $T_{F,q}$~\cite{Qiu:1991pp,  Qiu:1991wg, Qiu:1998ia}. Recently, the collinear twist-3 factorization has been verified at one-loop order, marking a significant achievement~\cite{Rein:2025qhe, Rein:2025pwu}.  A comprehensive study of the matching between the $k_\perp$-odd TMDs and the collinear twist-3 correlations has been carried out in Refs.~\cite{Zhou:2009jm, Rein:2022odl}.

The transverse-spin-dependent cross section for $W^\pm$ boson production in polarized $pp$ collisions can be written as~\cite{Kang:2011mr}
\begin{align}
    &\frac{\d \sigma_{UT}(S_\perp)}{ \d y \, \d^2 \vec{q}_\perp} =  \sin(\phi_q-\phi_S) \, \sigma_0 \int_0^\infty\frac{b^2 \, \d b}{4\pi} J_1 (b\,q_\perp) \notag \\
    &\quad \times \sum\limits_{q,q'} |V_{qq'}|^2 T_{F,q}(x_q,x_q,\mu_b)f_{q'}(x_{q'},\mu_b)e^{-S_{\rm P}(b)}\,.
\end{align}
In general, the SSA $A_{N}$ for $W^\pm$ boson production is given by
\begin{equation}\label{eq:SSA}
    A_{N} = -\frac{\int_{0}^{2\pi}\d\phi_q \,2 \sin(\phi_q-\phi_S) \, \d \sigma_{UT}}{\int_0^{2\pi}\d \phi_q \, \d \sigma_{UU}}\,.
\end{equation}
The SSA for $Z^0$ production takes a similar form,  with appropriate modifications for the electroweak couplings.

{\it Phenomenology} ---
\begin{figure}
\centering
 \includegraphics[scale=0.54]{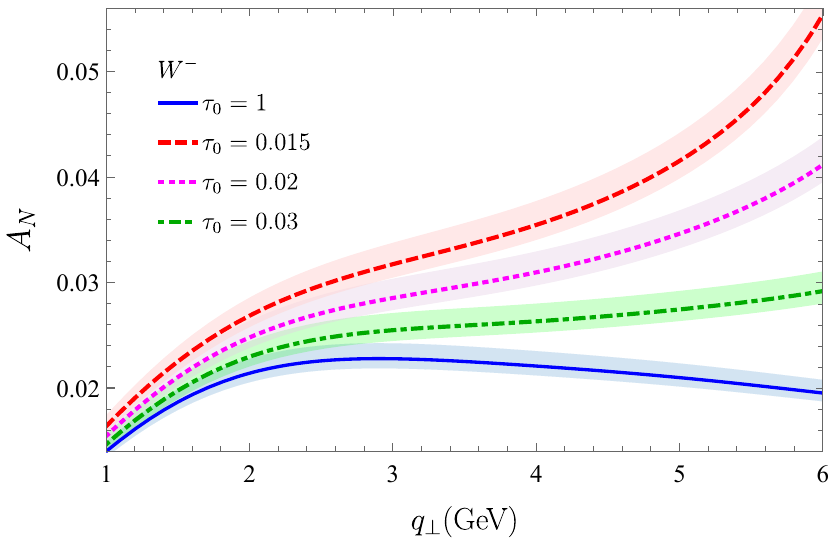}

 \includegraphics[scale=0.41]{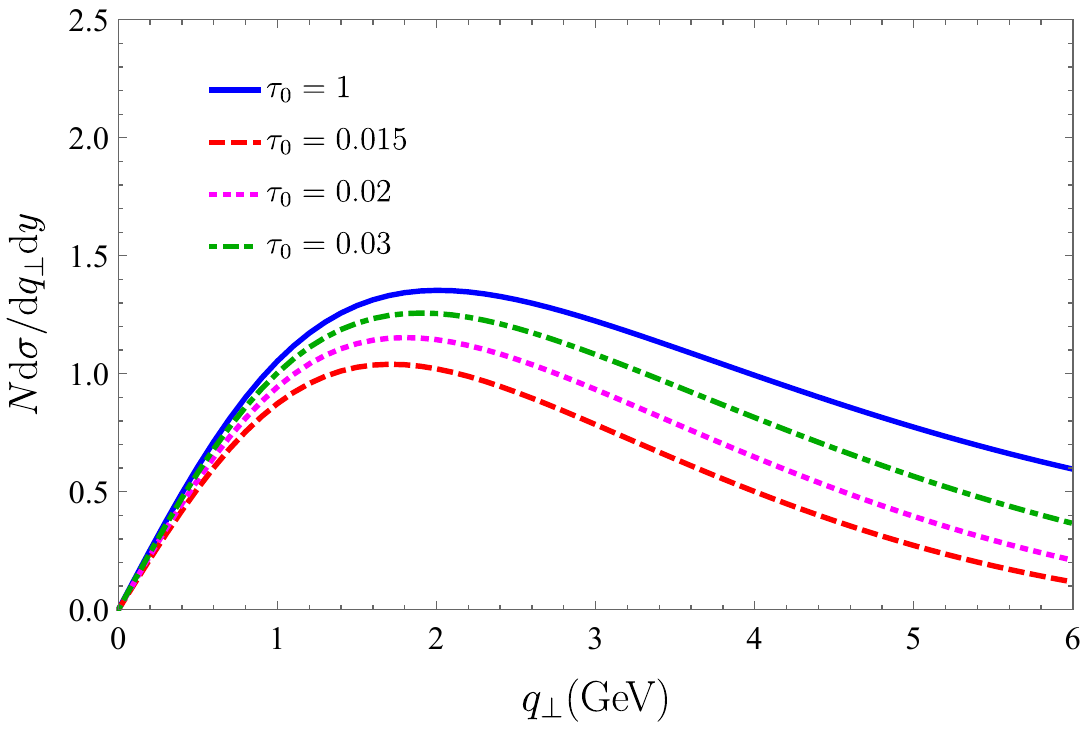}

 \caption{Top: SSA for $W^-$ production versus transverse momentum $q_\perp$. Bands correspond to theoretical uncertainties estimated via scale variation. Bottom: Corresponding normalized transverse momentum distribution. Results are for polarized $pp$ collisions at the RHIC energy $\sqrt{s}=500\,$GeV and the rapidity $y=0$, comparing various values of $\tau_0$ (indicated by color and style). }
\label{fig:Wm_th}
\end{figure}
We now present numerical results to demonstrate the implications of the key results in Eq.~\eqref{eq:SSA}. To ensure a smooth transition from the perturbative to the non-perturbative regions, the Sudakov factor is conventionally decomposed into two components,
\begin{eqnarray} 
S(b)=S_{\text P}(b_*)+S_{\text{NP}}(b)\,,
\end{eqnarray}
where the scale $\mu_b$ in the perturbative Sudakov factor $S_{\text P}(b_*)$ is replaced by $\mu_{b_*}\equiv 2\,e^{-\gamma_E}/b_{*}$ with $b_*=b/\sqrt{1+b^2/b^2_{\rm max}}$ and $b_{\rm max}=1.5\,$GeV$^{-1}$ \cite{Collins:1984kg}. Note that, in our numerical evaluation, the DGLAP evolution of PDFs is frozen for $\mu_b^2>t$ \cite{Stewart:2009yx}. The parametrizations for the non-perturbative component of the Sudakov factor associated with the unpolarized quark TMDs are taken from~\cite{Sun:2014dqm},
\begin{eqnarray} 
S_{\text{NP}}^f(b)&=&0.42 \ln \frac{Q}{Q_0} \ln \frac{b}{b_*}+0.106 \, b^2\,, 
\end{eqnarray}
with $Q_0=\sqrt{2.4}\,$GeV. For the Qiu-Sterman function $T_{F,q}$,  the most economical choice is to parameterize it in terms of the unpolarized collinear PDF~\cite{Sun:2013dya, Echevarria:2014xaa},
\begin{eqnarray} 
T_{F,q}(x,x,\mu)={\cal N}_q(x) f_q(x,\mu)\,,
\end{eqnarray}
where we adopt a parametrization for the function ${\cal N}_q(x) $ given in Table 2 in Ref.~\cite{Echevarria:2020hpy} (see also~\cite{Bacchetta:2020gko, Bury:2021sue} for the state-of-the-art extractions of the Sivers functions). It is worth noting that the diagonal component of the scale evolution equation for the Qiu-Sterman function closely resembles the standard DGLAP kernel for the normal unpolarized quark PDF~\cite{Kang:2008ey, Zhou:2008mz, Vogelsang:2009pj, Braun:2009mi, Ma:2011nd, Schafer:2012ra, Ma:2012xn, Kang:2012em, Sun:2013hua, Zhou:2015lxa, delRio:2024vvq}. Consequently, this parametrization effectively addresses the scale dependence of the Qiu-Sterman function.  In addition, we use the non-perturbative Sudakov factor for the Sivers distribution obtained from the same global fitting~\cite{Echevarria:2020hpy},
\begin{eqnarray} 
S_{\text{NP}}^s(b)=0.42 \ln \frac{Q}{Q_0} \ln \frac{b}{b_*}+0.18 \, b^2\,.
\end{eqnarray}
Moreover, the one-loop running coupling $\alpha_s(\mu)$ is used in our numerical estimations.

At RHIC energies, the SSAs for $W^+$ and $Z^0$ production are predicted to be small (see Supplemental Material \cite{supplemental}), so we focus on $W^-$ production here. In Fig.~\ref{fig:Wm_th}, we present theoretical predictions for SSA ($A_N$) versus $q_\perp$ (top) and normalized $q_\perp$ distribution (bottom) at $\sqrt{s}=500\,$GeV and $y=0$. Curves compare various $\tau_0$ values, with $\tau_0=1$ as inclusive. Central predictions use the hard scale $Q$, with uncertainties from varying only this scale between $Q/2$ and $2Q$ in Eq.~\eqref{eq:Sud_joint}. As expected, SSA magnitude increases as $\tau_0$ decreases, thus enhancing Sivers sensitivity. 

\begin{figure}
\centering
 \includegraphics[scale=0.8]{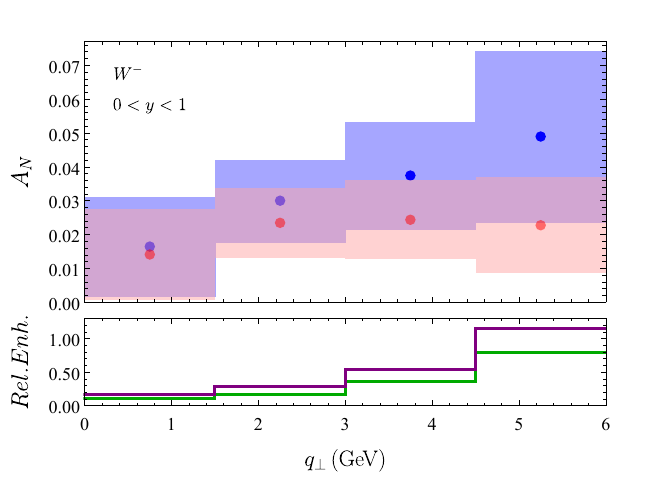}
 \includegraphics[scale=0.8]{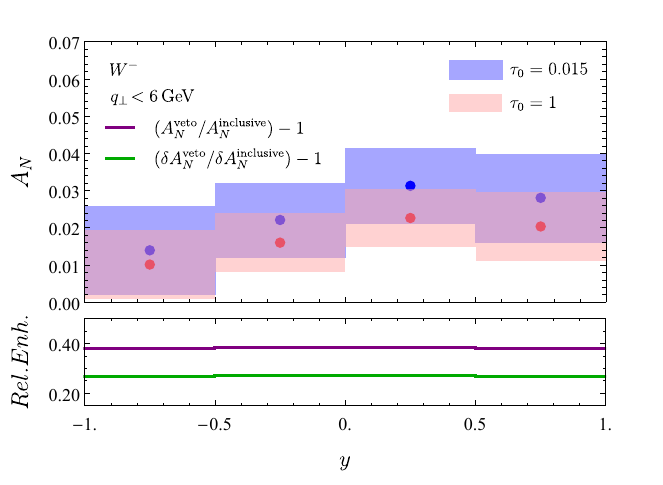}
 \caption{ Projected SSA for $W^-$ production in polarized $pp$ collisions at $\sqrt{s}=500\,$GeV.
Top half: Distributions versus transverse momentum $q_\perp$ (integrated over $0 < y < 1$). The upper panel shows $A_N$ for the inclusive case ($\tau_0=1$, red) and with a 0-jettiness veto ($\tau_0=0.015$, blue). Bands represent estimated statistical uncertainties (68\% CL). The lower panel shows the relative enhancement of the asymmetry (purple), compared to the relative increase in statistical uncertainty (green). Bottom half: Distributions versus rapidity $y$ (integrated over $q_\perp < 6$ GeV). Panels are as described for the top half.}
\label{fig:Wm_exp}
\end{figure}

The estimate of the experimental uncertainties, i.e., the statistical uncertainty of $A_{N}$, corresponding to a $68\%$ confidence level (CL) for a given integrated luminosity $\mathcal{L}$ and beam polarization rate $P$ is given by
\begin{eqnarray}\label{eq:StatUncert}
    \delta A_{N} =\frac{1}{P} \sqrt{\frac{1-(A_{N})^2}{\sigma \cdot \mathcal{L}}} \simeq \frac{1}{P}\frac{1}{\sqrt{\sigma \cdot \mathcal{L}}}\,,
\end{eqnarray}
where the second (approximated) equality holds due to the expected small value of $A_{N}$. For projected performance at RHIC, we assume polarized $pp$ collisions at the center-of-mass energy $\sqrt{s}=500\,{\rm GeV}$ with an integrated luminosity of $\mathcal{L} = 780\,{\rm pb}^{-1}$ and effective polarization $P=53\%$ \cite{footnote}.

Fig.~\ref{fig:Wm_exp} presents the projected $A_N$ for $W^-$ production, comparing the inclusive ($\tau_0=1$, red) and vetoed ($\tau_0=0.015$, blue) scenarios at $\sqrt{s}=500\,$GeV. The top half shows the distributions versus transverse momentum $q_\perp$ (integrated over $0 < y < 1$), and the bottom half shows the distributions versus rapidity $y$ (integrated over $q_\perp < 6$ GeV). In both halves, the upper panels plot the $A_N$ values, including estimated statistical error bands (68\% CL) based on Eq.~\eqref{eq:StatUncert}. A quantitative comparison in the $q_\perp$ distribution shows that the enhancement in the asymmetry is comparable to the estimated combined statistical uncertainty, indicating that the net gain is statistically significant and measurable at RHIC. To visualize this trade-off directly, the lower panels plot the relative enhancement of the asymmetry, $(A_N^{\text{veto}} / A_N^{\text{inclusive}}) - 1$ (purple line), against the relative increase in statistical uncertainty, $(\delta A_N^{\text{veto}} / \delta A_N^{\text{inclusive}}) - 1$ (green line). Across the entire kinematic range in both $q_\perp$ and $y$, the relative gain in the asymmetry (purple line) significantly exceeds the relative cost in precision (green line). For example, in the $q_\perp \approx 5$ GeV bin (top-right panel), the asymmetry increases by $\sim 115\%$, while the statistical uncertainty increases by only $\sim 80\%$. This confirms that the 0-jettiness veto yields a substantial net improvement in sensitivity. These results demonstrate that implementing a 0-jettiness veto offers a promising strategy to enhance the sensitivity of SSA measurements to the predicted sign flip of the Sivers function.

\begin{figure}
   \centering
 \includegraphics[scale=0.45]{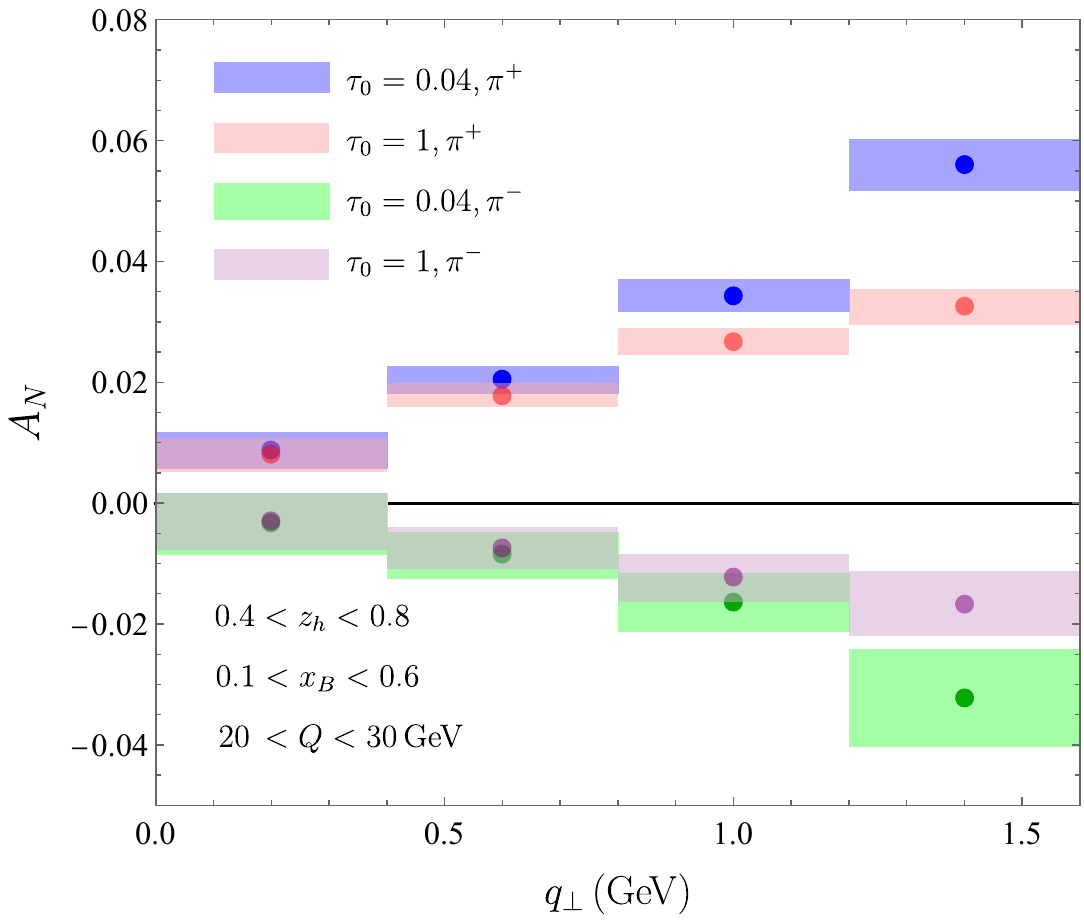}
     \caption{ The SSAs for $\pi^+$ (blue and red) and $\pi^-$ (green and purple) production in SIDIS at the EIC ($\sqrt{s} = 100\,$GeV) as a function of $q_\perp$. Predictions compare the inclusive case ($\tau_0=1$, red and purple bands) with a 1-jettiness veto ($\tau_0=0.04$, blue and green bands). Kinematic selections are $0.4<z_h<0.8$, $0.1<x_{B}<0.6$, and $20<Q<30\,$GeV. Statistical uncertainty bands (68\% CL) assume a projected integrated luminosity of $\mathcal{L}=100\,$fb$^{-1}$ and beam polarization $P=70\%$~\cite{AbdulKhalek:2021gbh, Nadel-Turonski:2025sfv}.}
 \label{plot3}
 \end{figure}

Our formalism can be naturally extended to describe the produced hadron's transverse momentum distribution in the SIDIS process. The relevant factorization framework associated with 1-jettiness in deep inelastic scattering (DIS) has been previously studied in the context of inclusive observables~\cite{Kang:2012zr, Kang:2013nha, Kang:2013lga, Cao:2024ota}.  In Fig.~\ref{plot3}, we apply our method to SSAs for $\pi^+$ and $\pi^-$ production at the future EIC, plotting the predictions with projected statistical uncertainty bands (68\% CL) based on Eq.~\eqref{eq:StatUncert} (adapted for EIC parameters). The results, with and without imposing a veto of $\tau_0 = 0.04$, demonstrate a clear enhancement in the asymmetries at moderate $q_\perp$ when the veto is applied. The statistical separation is clear: the error bands for the $\pi^+$ vetoed and inclusive cases show no overlap for $q_\perp \gtrsim 1$ GeV, confirming a measurable improvement in sensitivity.

{\it Conclusions and outlook} --- 
In this Letter, we establish a novel method for probing the nucleon's spin structure by employing a 0-jettiness veto to suppress TMD evolution effects. This approach directly addresses the long-standing challenge of disentangling non-perturbative dynamics from perturbative radiation. To this end, we derive a $k_t$-resummation formalism for 0-jettiness events and demonstrate its phenomenological impact by analyzing SSAs in $W^{\pm}/Z^0$ production at RHIC. We find that the veto leads to a remarkable enhancement of the asymmetry, enabling a more definitive test of the predicted sign change of the Sivers function.

This approach broadens the potential of TMD studies across a variety of experimental settings. The ability to systematically tune the veto scale, or to employ other veto observables like jet transverse momentum or vector angularities~\cite{Bijl:2023dux}, provides a powerful new lever to isolate intrinsic parton motion and constrain the non-perturbative Sudakov factor~\cite{Collins:2014jpa}. The framework's versatility extends to studying TMDs in di-jet and electron-jet production at the EIC, while offering particular advantages for TMD measurements at the LHC, where evolution effects are more pronounced. These developments pave the way for a new era of precision in mapping the nucleon's 3D structure.

\

{\it Acknowledgments:} We thank Qing-hua Xu for helpful discussions. This work has been supported by the National Natural Science Foundation of China under Grants No.~12175118 (J.~Z.), No.~12321005 (J.~Z.), No.~12275052 (S.~F. and D.~Y.~S.)  and No.~12147101 (S.~F. and D.~Y.~S.). D.~Y.~S. is also supported by the Innovation Program for Quantum Science and Technology under Grant No.~2024ZD0300101.
\

\bibliography{ref.bib}

\begin{thebibliography}{91}
\expandafter\ifx\csname natexlab\endcsname\relax\def\natexlab#1{#1}\fi
\expandafter\ifx\csname bibnamefont\endcsname\relax
  \def\bibnamefont#1{#1}\fi
\expandafter\ifx\csname bibfnamefont\endcsname\relax
  \def\bibfnamefont#1{#1}\fi
\expandafter\ifx\csname citenamefont\endcsname\relax
  \def\citenamefont#1{#1}\fi
\expandafter\ifx\csname url\endcsname\relax
  \def\url#1{\texttt{#1}}\fi
\expandafter\ifx\csname urlprefix\endcsname\relax\def\urlprefix{URL }\fi
\providecommand{\bibinfo}[2]{#2}
\providecommand{\eprint}[2][]{\url{#2}}

\bibitem[{\citenamefont{Accardi et~al.}(2016)}]{Accardi:2012qut}
\bibinfo{author}{\bibfnamefont{A.}~\bibnamefont{Accardi}} \bibnamefont{et~al.}, \bibinfo{journal}{Eur. Phys. J. A} \textbf{\bibinfo{volume}{52}}, \bibinfo{pages}{268} (\bibinfo{year}{2016}), \eprint{1212.1701}.

\bibitem[{\citenamefont{Abdul~Khalek et~al.}(2022)}]{AbdulKhalek:2021gbh}
\bibinfo{author}{\bibfnamefont{R.}~\bibnamefont{Abdul~Khalek}} \bibnamefont{et~al.}, \bibinfo{journal}{Nucl. Phys. A} \textbf{\bibinfo{volume}{1026}}, \bibinfo{pages}{122447} (\bibinfo{year}{2022}), \eprint{2103.05419}.

\bibitem[{\citenamefont{Anderle et~al.}(2021)}]{Anderle:2021wcy}
\bibinfo{author}{\bibfnamefont{D.~P.} \bibnamefont{Anderle}} \bibnamefont{et~al.}, \bibinfo{journal}{Front. Phys. (Beijing)} \textbf{\bibinfo{volume}{16}}, \bibinfo{pages}{64701} (\bibinfo{year}{2021}), \eprint{2102.09222}.

\bibitem[{\citenamefont{Collins and Soper}(1981)}]{Collins:1981uk}
\bibinfo{author}{\bibfnamefont{J.~C.} \bibnamefont{Collins}} \bibnamefont{and} \bibinfo{author}{\bibfnamefont{D.~E.} \bibnamefont{Soper}}, \bibinfo{journal}{Nucl. Phys. B} \textbf{\bibinfo{volume}{193}}, \bibinfo{pages}{381} (\bibinfo{year}{1981}), \bibinfo{note}{[Erratum: Nucl.Phys.B 213, 545 (1983)]}.

\bibitem[{\citenamefont{Collins and Soper}(1982)}]{Collins:1981uw}
\bibinfo{author}{\bibfnamefont{J.~C.} \bibnamefont{Collins}} \bibnamefont{and} \bibinfo{author}{\bibfnamefont{D.~E.} \bibnamefont{Soper}}, \bibinfo{journal}{Nucl. Phys. B} \textbf{\bibinfo{volume}{194}}, \bibinfo{pages}{445} (\bibinfo{year}{1982}).

\bibitem[{\citenamefont{Ji et~al.}(2005)\citenamefont{Ji, Ma, and Yuan}}]{Ji:2004wu}
\bibinfo{author}{\bibfnamefont{X.-d.} \bibnamefont{Ji}}, \bibinfo{author}{\bibfnamefont{J.-p.} \bibnamefont{Ma}}, \bibnamefont{and} \bibinfo{author}{\bibfnamefont{F.}~\bibnamefont{Yuan}}, \bibinfo{journal}{Phys. Rev. D} \textbf{\bibinfo{volume}{71}}, \bibinfo{pages}{034005} (\bibinfo{year}{2005}), \eprint{hep-ph/0404183}.

\bibitem[{\citenamefont{Collins}(2011)}]{Collins:2011zzd}
\bibinfo{author}{\bibfnamefont{J.}~\bibnamefont{Collins}}, \emph{\bibinfo{title}{{Foundations of Perturbative QCD}}}, vol.~\bibinfo{volume}{32} (\bibinfo{publisher}{Cambridge University Press}, \bibinfo{year}{2011}), ISBN \bibinfo{isbn}{978-1-009-40184-5, 978-1-009-40183-8, 978-1-009-40182-1}.

\bibitem[{\citenamefont{Boussarie et~al.}(2023)}]{Boussarie:2023izj}
\bibinfo{author}{\bibfnamefont{R.}~\bibnamefont{Boussarie}} \bibnamefont{et~al.} (\bibinfo{year}{2023}), \eprint{2304.03302}.

\bibitem[{\citenamefont{Mulders and Tangerman}(1996)}]{Mulders:1995dh}
\bibinfo{author}{\bibfnamefont{P.~J.} \bibnamefont{Mulders}} \bibnamefont{and} \bibinfo{author}{\bibfnamefont{R.~D.} \bibnamefont{Tangerman}}, \bibinfo{journal}{Nucl. Phys. B} \textbf{\bibinfo{volume}{461}}, \bibinfo{pages}{197} (\bibinfo{year}{1996}), \bibinfo{note}{[Erratum: Nucl.Phys.B 484, 538--540 (1997)]}, \eprint{hep-ph/9510301}.

\bibitem[{\citenamefont{Mulders and Rodrigues}(2001)}]{Mulders:2000sh}
\bibinfo{author}{\bibfnamefont{P.~J.} \bibnamefont{Mulders}} \bibnamefont{and} \bibinfo{author}{\bibfnamefont{J.}~\bibnamefont{Rodrigues}}, \bibinfo{journal}{Phys. Rev. D} \textbf{\bibinfo{volume}{63}}, \bibinfo{pages}{094021} (\bibinfo{year}{2001}), \eprint{hep-ph/0009343}.

\bibitem[{\citenamefont{Sivers}(1990)}]{Sivers:1989cc}
\bibinfo{author}{\bibfnamefont{D.~W.} \bibnamefont{Sivers}}, \bibinfo{journal}{Phys. Rev. D} \textbf{\bibinfo{volume}{41}}, \bibinfo{pages}{83} (\bibinfo{year}{1990}).

\bibitem[{\citenamefont{Collins}(1993)}]{Collins:1992kk}
\bibinfo{author}{\bibfnamefont{J.~C.} \bibnamefont{Collins}}, \bibinfo{journal}{Nucl. Phys. B} \textbf{\bibinfo{volume}{396}}, \bibinfo{pages}{161} (\bibinfo{year}{1993}), \eprint{hep-ph/9208213}.

\bibitem[{\citenamefont{Liang and Boros}(1997)}]{Liang:1997rt}
\bibinfo{author}{\bibfnamefont{Z.-t.} \bibnamefont{Liang}} \bibnamefont{and} \bibinfo{author}{\bibfnamefont{C.}~\bibnamefont{Boros}}, \bibinfo{journal}{Phys. Rev. Lett.} \textbf{\bibinfo{volume}{79}}, \bibinfo{pages}{3608} (\bibinfo{year}{1997}), \eprint{hep-ph/9708488}.

\bibitem[{\citenamefont{Ji et~al.}(2003)\citenamefont{Ji, Ma, and Yuan}}]{Ji:2002xn}
\bibinfo{author}{\bibfnamefont{X.-d.} \bibnamefont{Ji}}, \bibinfo{author}{\bibfnamefont{J.-P.} \bibnamefont{Ma}}, \bibnamefont{and} \bibinfo{author}{\bibfnamefont{F.}~\bibnamefont{Yuan}}, \bibinfo{journal}{Nucl. Phys. B} \textbf{\bibinfo{volume}{652}}, \bibinfo{pages}{383} (\bibinfo{year}{2003}), \eprint{hep-ph/0210430}.

\bibitem[{\citenamefont{Brodsky et~al.}(2002{\natexlab{a}})\citenamefont{Brodsky, Hwang, and Schmidt}}]{Brodsky:2002cx}
\bibinfo{author}{\bibfnamefont{S.~J.} \bibnamefont{Brodsky}}, \bibinfo{author}{\bibfnamefont{D.~S.} \bibnamefont{Hwang}}, \bibnamefont{and} \bibinfo{author}{\bibfnamefont{I.}~\bibnamefont{Schmidt}}, \bibinfo{journal}{Phys. Lett. B} \textbf{\bibinfo{volume}{530}}, \bibinfo{pages}{99} (\bibinfo{year}{2002}{\natexlab{a}}), \eprint{hep-ph/0201296}.

\bibitem[{\citenamefont{Brodsky et~al.}(2002{\natexlab{b}})\citenamefont{Brodsky, Hwang, and Schmidt}}]{Brodsky:2002rv}
\bibinfo{author}{\bibfnamefont{S.~J.} \bibnamefont{Brodsky}}, \bibinfo{author}{\bibfnamefont{D.~S.} \bibnamefont{Hwang}}, \bibnamefont{and} \bibinfo{author}{\bibfnamefont{I.}~\bibnamefont{Schmidt}}, \bibinfo{journal}{Nucl. Phys. B} \textbf{\bibinfo{volume}{642}}, \bibinfo{pages}{344} (\bibinfo{year}{2002}{\natexlab{b}}), \eprint{hep-ph/0206259}.

\bibitem[{\citenamefont{Collins}(2002)}]{Collins:2002kn}
\bibinfo{author}{\bibfnamefont{J.~C.} \bibnamefont{Collins}}, \bibinfo{journal}{Phys. Lett. B} \textbf{\bibinfo{volume}{536}}, \bibinfo{pages}{43} (\bibinfo{year}{2002}), \eprint{hep-ph/0204004}.

\bibitem[{\citenamefont{Collins and Metz}(2004)}]{Collins:2004nx}
\bibinfo{author}{\bibfnamefont{J.~C.} \bibnamefont{Collins}} \bibnamefont{and} \bibinfo{author}{\bibfnamefont{A.}~\bibnamefont{Metz}}, \bibinfo{journal}{Phys. Rev. Lett.} \textbf{\bibinfo{volume}{93}}, \bibinfo{pages}{252001} (\bibinfo{year}{2004}), \eprint{hep-ph/0408249}.

\bibitem[{\citenamefont{Ji and Yuan}(2002)}]{Ji:2002aa}
\bibinfo{author}{\bibfnamefont{X.-d.} \bibnamefont{Ji}} \bibnamefont{and} \bibinfo{author}{\bibfnamefont{F.}~\bibnamefont{Yuan}}, \bibinfo{journal}{Phys. Lett. B} \textbf{\bibinfo{volume}{543}}, \bibinfo{pages}{66} (\bibinfo{year}{2002}), \eprint{hep-ph/0206057}.

\bibitem[{\citenamefont{McLerran and Venugopalan}(1994{\natexlab{a}})}]{McLerran:1993ni}
\bibinfo{author}{\bibfnamefont{L.~D.} \bibnamefont{McLerran}} \bibnamefont{and} \bibinfo{author}{\bibfnamefont{R.}~\bibnamefont{Venugopalan}}, \bibinfo{journal}{Phys. Rev. D} \textbf{\bibinfo{volume}{49}}, \bibinfo{pages}{2233} (\bibinfo{year}{1994}{\natexlab{a}}), \eprint{hep-ph/9309289}.

\bibitem[{\citenamefont{McLerran and Venugopalan}(1994{\natexlab{b}})}]{McLerran:1993ka}
\bibinfo{author}{\bibfnamefont{L.~D.} \bibnamefont{McLerran}} \bibnamefont{and} \bibinfo{author}{\bibfnamefont{R.}~\bibnamefont{Venugopalan}}, \bibinfo{journal}{Phys. Rev. D} \textbf{\bibinfo{volume}{49}}, \bibinfo{pages}{3352} (\bibinfo{year}{1994}{\natexlab{b}}), \eprint{hep-ph/9311205}.

\bibitem[{\citenamefont{Dominguez et~al.}(2011)\citenamefont{Dominguez, Marquet, Xiao, and Yuan}}]{Dominguez:2011wm}
\bibinfo{author}{\bibfnamefont{F.}~\bibnamefont{Dominguez}}, \bibinfo{author}{\bibfnamefont{C.}~\bibnamefont{Marquet}}, \bibinfo{author}{\bibfnamefont{B.-W.} \bibnamefont{Xiao}}, \bibnamefont{and} \bibinfo{author}{\bibfnamefont{F.}~\bibnamefont{Yuan}}, \bibinfo{journal}{Phys. Rev. D} \textbf{\bibinfo{volume}{83}}, \bibinfo{pages}{105005} (\bibinfo{year}{2011}), \eprint{1101.0715}.

\bibitem[{\citenamefont{Metz and Zhou}(2011{\natexlab{a}})}]{Metz:2011wb}
\bibinfo{author}{\bibfnamefont{A.}~\bibnamefont{Metz}} \bibnamefont{and} \bibinfo{author}{\bibfnamefont{J.}~\bibnamefont{Zhou}}, \bibinfo{journal}{Phys. Rev. D} \textbf{\bibinfo{volume}{84}}, \bibinfo{pages}{051503} (\bibinfo{year}{2011}{\natexlab{a}}), \eprint{1105.1991}.

\bibitem[{\citenamefont{Akcakaya et~al.}(2013)\citenamefont{Akcakaya, Sch\"afer, and Zhou}}]{Akcakaya:2012si}
\bibinfo{author}{\bibfnamefont{E.}~\bibnamefont{Akcakaya}}, \bibinfo{author}{\bibfnamefont{A.}~\bibnamefont{Sch\"afer}}, \bibnamefont{and} \bibinfo{author}{\bibfnamefont{J.}~\bibnamefont{Zhou}}, \bibinfo{journal}{Phys. Rev. D} \textbf{\bibinfo{volume}{87}}, \bibinfo{pages}{054010} (\bibinfo{year}{2013}), \eprint{1208.4965}.

\bibitem[{\citenamefont{Zhou}(2014)}]{Zhou:2013gsa}
\bibinfo{author}{\bibfnamefont{J.}~\bibnamefont{Zhou}}, \bibinfo{journal}{Phys. Rev. D} \textbf{\bibinfo{volume}{89}}, \bibinfo{pages}{074050} (\bibinfo{year}{2014}), \eprint{1308.5912}.

\bibitem[{\citenamefont{Kotko et~al.}(2015)\citenamefont{Kotko, Kutak, Marquet, Petreska, Sapeta, and van Hameren}}]{Kotko:2015ura}
\bibinfo{author}{\bibfnamefont{P.}~\bibnamefont{Kotko}}, \bibinfo{author}{\bibfnamefont{K.}~\bibnamefont{Kutak}}, \bibinfo{author}{\bibfnamefont{C.}~\bibnamefont{Marquet}}, \bibinfo{author}{\bibfnamefont{E.}~\bibnamefont{Petreska}}, \bibinfo{author}{\bibfnamefont{S.}~\bibnamefont{Sapeta}}, \bibnamefont{and} \bibinfo{author}{\bibfnamefont{A.}~\bibnamefont{van Hameren}}, \bibinfo{journal}{JHEP} \textbf{\bibinfo{volume}{09}}, \bibinfo{pages}{106} (\bibinfo{year}{2015}), \eprint{1503.03421}.

\bibitem[{\citenamefont{Boer et~al.}(2016)\citenamefont{Boer, Echevarria, Mulders, and Zhou}}]{Boer:2015pni}
\bibinfo{author}{\bibfnamefont{D.}~\bibnamefont{Boer}}, \bibinfo{author}{\bibfnamefont{M.~G.} \bibnamefont{Echevarria}}, \bibinfo{author}{\bibfnamefont{P.}~\bibnamefont{Mulders}}, \bibnamefont{and} \bibinfo{author}{\bibfnamefont{J.}~\bibnamefont{Zhou}}, \bibinfo{journal}{Phys. Rev. Lett.} \textbf{\bibinfo{volume}{116}}, \bibinfo{pages}{122001} (\bibinfo{year}{2016}), \eprint{1511.03485}.

\bibitem[{\citenamefont{Balitsky and Tarasov}(2016)}]{Balitsky:2016dgz}
\bibinfo{author}{\bibfnamefont{I.}~\bibnamefont{Balitsky}} \bibnamefont{and} \bibinfo{author}{\bibfnamefont{A.}~\bibnamefont{Tarasov}}, \bibinfo{journal}{JHEP} \textbf{\bibinfo{volume}{06}}, \bibinfo{pages}{164} (\bibinfo{year}{2016}), \eprint{1603.06548}.

\bibitem[{\citenamefont{Altinoluk and Boussarie}(2019)}]{Altinoluk:2019wyu}
\bibinfo{author}{\bibfnamefont{T.}~\bibnamefont{Altinoluk}} \bibnamefont{and} \bibinfo{author}{\bibfnamefont{R.}~\bibnamefont{Boussarie}}, \bibinfo{journal}{JHEP} \textbf{\bibinfo{volume}{10}}, \bibinfo{pages}{208} (\bibinfo{year}{2019}), \eprint{1902.07930}.

\bibitem[{\citenamefont{Stewart et~al.}(2010{\natexlab{a}})\citenamefont{Stewart, Tackmann, and Waalewijn}}]{Stewart:2010tn}
\bibinfo{author}{\bibfnamefont{I.~W.} \bibnamefont{Stewart}}, \bibinfo{author}{\bibfnamefont{F.~J.} \bibnamefont{Tackmann}}, \bibnamefont{and} \bibinfo{author}{\bibfnamefont{W.~J.} \bibnamefont{Waalewijn}}, \bibinfo{journal}{Phys. Rev. Lett.} \textbf{\bibinfo{volume}{105}}, \bibinfo{pages}{092002} (\bibinfo{year}{2010}{\natexlab{a}}), \eprint{1004.2489}.

\bibitem[{\citenamefont{Gaunt et~al.}(2015)\citenamefont{Gaunt, Stahlhofen, Tackmann, and Walsh}}]{Gaunt:2015pea}
\bibinfo{author}{\bibfnamefont{J.}~\bibnamefont{Gaunt}}, \bibinfo{author}{\bibfnamefont{M.}~\bibnamefont{Stahlhofen}}, \bibinfo{author}{\bibfnamefont{F.~J.} \bibnamefont{Tackmann}}, \bibnamefont{and} \bibinfo{author}{\bibfnamefont{J.~R.} \bibnamefont{Walsh}}, \bibinfo{journal}{JHEP} \textbf{\bibinfo{volume}{09}}, \bibinfo{pages}{058} (\bibinfo{year}{2015}), \eprint{1505.04794}.

\bibitem[{\citenamefont{Boughezal et~al.}(2015{\natexlab{a}})\citenamefont{Boughezal, Focke, Giele, Liu, and Petriello}}]{Boughezal:2015aha}
\bibinfo{author}{\bibfnamefont{R.}~\bibnamefont{Boughezal}}, \bibinfo{author}{\bibfnamefont{C.}~\bibnamefont{Focke}}, \bibinfo{author}{\bibfnamefont{W.}~\bibnamefont{Giele}}, \bibinfo{author}{\bibfnamefont{X.}~\bibnamefont{Liu}}, \bibnamefont{and} \bibinfo{author}{\bibfnamefont{F.}~\bibnamefont{Petriello}}, \bibinfo{journal}{Phys. Lett. B} \textbf{\bibinfo{volume}{748}}, \bibinfo{pages}{5} (\bibinfo{year}{2015}{\natexlab{a}}), \eprint{1505.03893}.

\bibitem[{\citenamefont{Boughezal et~al.}(2015{\natexlab{b}})\citenamefont{Boughezal, Focke, Liu, and Petriello}}]{Boughezal:2015dva}
\bibinfo{author}{\bibfnamefont{R.}~\bibnamefont{Boughezal}}, \bibinfo{author}{\bibfnamefont{C.}~\bibnamefont{Focke}}, \bibinfo{author}{\bibfnamefont{X.}~\bibnamefont{Liu}}, \bibnamefont{and} \bibinfo{author}{\bibfnamefont{F.}~\bibnamefont{Petriello}}, \bibinfo{journal}{Phys. Rev. Lett.} \textbf{\bibinfo{volume}{115}}, \bibinfo{pages}{062002} (\bibinfo{year}{2015}{\natexlab{b}}), \eprint{1504.02131}.

\bibitem[{\citenamefont{Boughezal et~al.}(2017)\citenamefont{Boughezal, Campbell, Ellis, Focke, Giele, Liu, Petriello, and Williams}}]{Boughezal:2016wmq}
\bibinfo{author}{\bibfnamefont{R.}~\bibnamefont{Boughezal}}, \bibinfo{author}{\bibfnamefont{J.~M.} \bibnamefont{Campbell}}, \bibinfo{author}{\bibfnamefont{R.~K.} \bibnamefont{Ellis}}, \bibinfo{author}{\bibfnamefont{C.}~\bibnamefont{Focke}}, \bibinfo{author}{\bibfnamefont{W.}~\bibnamefont{Giele}}, \bibinfo{author}{\bibfnamefont{X.}~\bibnamefont{Liu}}, \bibinfo{author}{\bibfnamefont{F.}~\bibnamefont{Petriello}}, \bibnamefont{and} \bibinfo{author}{\bibfnamefont{C.}~\bibnamefont{Williams}}, \bibinfo{journal}{Eur. Phys. J. C} \textbf{\bibinfo{volume}{77}}, \bibinfo{pages}{7} (\bibinfo{year}{2017}), \eprint{1605.08011}.

\bibitem[{\citenamefont{Stewart et~al.}(2010{\natexlab{b}})\citenamefont{Stewart, Tackmann, and Waalewijn}}]{Stewart:2010pd}
\bibinfo{author}{\bibfnamefont{I.~W.} \bibnamefont{Stewart}}, \bibinfo{author}{\bibfnamefont{F.~J.} \bibnamefont{Tackmann}}, \bibnamefont{and} \bibinfo{author}{\bibfnamefont{W.~J.} \bibnamefont{Waalewijn}}, \bibinfo{journal}{JHEP} \textbf{\bibinfo{volume}{09}}, \bibinfo{pages}{005} (\bibinfo{year}{2010}{\natexlab{b}}), \eprint{1002.2213}.

\bibitem[{\citenamefont{Kang et~al.}(2014)\citenamefont{Kang, Liu, and Mantry}}]{Kang:2013lga}
\bibinfo{author}{\bibfnamefont{Z.-B.} \bibnamefont{Kang}}, \bibinfo{author}{\bibfnamefont{X.}~\bibnamefont{Liu}}, \bibnamefont{and} \bibinfo{author}{\bibfnamefont{S.}~\bibnamefont{Mantry}}, \bibinfo{journal}{Phys. Rev. D} \textbf{\bibinfo{volume}{90}}, \bibinfo{pages}{014041} (\bibinfo{year}{2014}), \eprint{1312.0301}.

\bibitem[{\citenamefont{Lustermans et~al.}(2019)\citenamefont{Lustermans, Michel, Tackmann, and Waalewijn}}]{Lustermans:2019plv}
\bibinfo{author}{\bibfnamefont{G.}~\bibnamefont{Lustermans}}, \bibinfo{author}{\bibfnamefont{J.~K.~L.} \bibnamefont{Michel}}, \bibinfo{author}{\bibfnamefont{F.~J.} \bibnamefont{Tackmann}}, \bibnamefont{and} \bibinfo{author}{\bibfnamefont{W.~J.} \bibnamefont{Waalewijn}}, \bibinfo{journal}{JHEP} \textbf{\bibinfo{volume}{03}}, \bibinfo{pages}{124} (\bibinfo{year}{2019}), \eprint{1901.03331}.

\bibitem[{\citenamefont{Alioli et~al.}(2022)\citenamefont{Alioli, Broggio, and Lim}}]{Alioli:2021ggd}
\bibinfo{author}{\bibfnamefont{S.}~\bibnamefont{Alioli}}, \bibinfo{author}{\bibfnamefont{A.}~\bibnamefont{Broggio}}, \bibnamefont{and} \bibinfo{author}{\bibfnamefont{M.~A.} \bibnamefont{Lim}}, \bibinfo{journal}{JHEP} \textbf{\bibinfo{volume}{01}}, \bibinfo{pages}{066} (\bibinfo{year}{2022}), \eprint{2111.03632}.

\bibitem[{\citenamefont{Alioli et~al.}(2024)\citenamefont{Alioli, Bell, Billis, Broggio, Dehnadi, Lim, Marinelli, Nagar, Napoletano, and Rahn}}]{Alioli:2023rxx}
\bibinfo{author}{\bibfnamefont{S.}~\bibnamefont{Alioli}}, \bibinfo{author}{\bibfnamefont{G.}~\bibnamefont{Bell}}, \bibinfo{author}{\bibfnamefont{G.}~\bibnamefont{Billis}}, \bibinfo{author}{\bibfnamefont{A.}~\bibnamefont{Broggio}}, \bibinfo{author}{\bibfnamefont{B.}~\bibnamefont{Dehnadi}}, \bibinfo{author}{\bibfnamefont{M.~A.} \bibnamefont{Lim}}, \bibinfo{author}{\bibfnamefont{G.}~\bibnamefont{Marinelli}}, \bibinfo{author}{\bibfnamefont{R.}~\bibnamefont{Nagar}}, \bibinfo{author}{\bibfnamefont{D.}~\bibnamefont{Napoletano}}, \bibnamefont{and} \bibinfo{author}{\bibfnamefont{R.}~\bibnamefont{Rahn}}, \bibinfo{journal}{Phys. Rev. D} \textbf{\bibinfo{volume}{109}}, \bibinfo{pages}{094009} (\bibinfo{year}{2024}), \eprint{2312.06496}.

\bibitem[{\citenamefont{Knobbe et~al.}(2023)\citenamefont{Knobbe, Reichelt, and Schumann}}]{Knobbe:2023ehi}
\bibinfo{author}{\bibfnamefont{M.}~\bibnamefont{Knobbe}}, \bibinfo{author}{\bibfnamefont{D.}~\bibnamefont{Reichelt}}, \bibnamefont{and} \bibinfo{author}{\bibfnamefont{S.}~\bibnamefont{Schumann}}, \bibinfo{journal}{JHEP} \textbf{\bibinfo{volume}{09}}, \bibinfo{pages}{194} (\bibinfo{year}{2023}), \eprint{2306.17736}.

\bibitem[{\citenamefont{Cao et~al.}(2024)\citenamefont{Cao, Kang, Liu, and Mantry}}]{Cao:2024ota}
\bibinfo{author}{\bibfnamefont{H.}~\bibnamefont{Cao}}, \bibinfo{author}{\bibfnamefont{Z.-B.} \bibnamefont{Kang}}, \bibinfo{author}{\bibfnamefont{X.}~\bibnamefont{Liu}}, \bibnamefont{and} \bibinfo{author}{\bibfnamefont{S.}~\bibnamefont{Mantry}}, \bibinfo{journal}{Phys. Rev. D} \textbf{\bibinfo{volume}{110}}, \bibinfo{pages}{014045} (\bibinfo{year}{2024}), \eprint{2401.01941}.

\bibitem[{\citenamefont{Jain et~al.}(2012)\citenamefont{Jain, Procura, and Waalewijn}}]{Jain:2011iu}
\bibinfo{author}{\bibfnamefont{A.}~\bibnamefont{Jain}}, \bibinfo{author}{\bibfnamefont{M.}~\bibnamefont{Procura}}, \bibnamefont{and} \bibinfo{author}{\bibfnamefont{W.~J.} \bibnamefont{Waalewijn}}, \bibinfo{journal}{JHEP} \textbf{\bibinfo{volume}{04}}, \bibinfo{pages}{132} (\bibinfo{year}{2012}), \eprint{1110.0839}.

\bibitem[{\citenamefont{Procura et~al.}(2015)\citenamefont{Procura, Waalewijn, and Zeune}}]{Procura:2014cba}
\bibinfo{author}{\bibfnamefont{M.}~\bibnamefont{Procura}}, \bibinfo{author}{\bibfnamefont{W.~J.} \bibnamefont{Waalewijn}}, \bibnamefont{and} \bibinfo{author}{\bibfnamefont{L.}~\bibnamefont{Zeune}}, \bibinfo{journal}{JHEP} \textbf{\bibinfo{volume}{02}}, \bibinfo{pages}{117} (\bibinfo{year}{2015}), \eprint{1410.6483}.

\bibitem[{\citenamefont{Monni et~al.}(2020)\citenamefont{Monni, Rottoli, and Torrielli}}]{Monni:2019yyr}
\bibinfo{author}{\bibfnamefont{P.~F.} \bibnamefont{Monni}}, \bibinfo{author}{\bibfnamefont{L.}~\bibnamefont{Rottoli}}, \bibnamefont{and} \bibinfo{author}{\bibfnamefont{P.}~\bibnamefont{Torrielli}}, \bibinfo{journal}{Phys. Rev. Lett.} \textbf{\bibinfo{volume}{124}}, \bibinfo{pages}{252001} (\bibinfo{year}{2020}), \eprint{1909.04704}.

\bibitem[{\citenamefont{Makris et~al.}(2021)\citenamefont{Makris, Ringer, and Waalewijn}}]{Makris:2020ltr}
\bibinfo{author}{\bibfnamefont{Y.}~\bibnamefont{Makris}}, \bibinfo{author}{\bibfnamefont{F.}~\bibnamefont{Ringer}}, \bibnamefont{and} \bibinfo{author}{\bibfnamefont{W.~J.} \bibnamefont{Waalewijn}}, \bibinfo{journal}{JHEP} \textbf{\bibinfo{volume}{02}}, \bibinfo{pages}{070} (\bibinfo{year}{2021}), \eprint{2009.11871}.

\bibitem[{\citenamefont{Kang and Qiu}(2009{\natexlab{a}})}]{Kang:2009bp}
\bibinfo{author}{\bibfnamefont{Z.-B.} \bibnamefont{Kang}} \bibnamefont{and} \bibinfo{author}{\bibfnamefont{J.-W.} \bibnamefont{Qiu}}, \bibinfo{journal}{Phys. Rev. Lett.} \textbf{\bibinfo{volume}{103}}, \bibinfo{pages}{172001} (\bibinfo{year}{2009}{\natexlab{a}}), \eprint{0903.3629}.

\bibitem[{\citenamefont{Metz and Zhou}(2011{\natexlab{b}})}]{Metz:2010xs}
\bibinfo{author}{\bibfnamefont{A.}~\bibnamefont{Metz}} \bibnamefont{and} \bibinfo{author}{\bibfnamefont{J.}~\bibnamefont{Zhou}}, \bibinfo{journal}{Phys. Lett. B} \textbf{\bibinfo{volume}{700}}, \bibinfo{pages}{11} (\bibinfo{year}{2011}{\natexlab{b}}), \eprint{1006.3097}.

\bibitem[{\citenamefont{Kang et~al.}(2011)\citenamefont{Kang, Xiao, and Yuan}}]{Kang:2011mr}
\bibinfo{author}{\bibfnamefont{Z.-B.} \bibnamefont{Kang}}, \bibinfo{author}{\bibfnamefont{B.-W.} \bibnamefont{Xiao}}, \bibnamefont{and} \bibinfo{author}{\bibfnamefont{F.}~\bibnamefont{Yuan}}, \bibinfo{journal}{Phys. Rev. Lett.} \textbf{\bibinfo{volume}{107}}, \bibinfo{pages}{152002} (\bibinfo{year}{2011}), \eprint{1106.0266}.

\bibitem[{\citenamefont{Echevarria et~al.}(2021)\citenamefont{Echevarria, Kang, and Terry}}]{Echevarria:2020hpy}
\bibinfo{author}{\bibfnamefont{M.~G.} \bibnamefont{Echevarria}}, \bibinfo{author}{\bibfnamefont{Z.-B.} \bibnamefont{Kang}}, \bibnamefont{and} \bibinfo{author}{\bibfnamefont{J.}~\bibnamefont{Terry}}, \bibinfo{journal}{JHEP} \textbf{\bibinfo{volume}{01}}, \bibinfo{pages}{126} (\bibinfo{year}{2021}), \eprint{2009.10710}.

\bibitem[{\citenamefont{Qiu and Sterman}(1991)}]{Qiu:1991pp}
\bibinfo{author}{\bibfnamefont{J.-w.} \bibnamefont{Qiu}} \bibnamefont{and} \bibinfo{author}{\bibfnamefont{G.~F.} \bibnamefont{Sterman}}, \bibinfo{journal}{Phys. Rev. Lett.} \textbf{\bibinfo{volume}{67}}, \bibinfo{pages}{2264} (\bibinfo{year}{1991}).

\bibitem[{\citenamefont{Ji et~al.}(2006)\citenamefont{Ji, Qiu, Vogelsang, and Yuan}}]{Ji:2006ub}
\bibinfo{author}{\bibfnamefont{X.}~\bibnamefont{Ji}}, \bibinfo{author}{\bibfnamefont{J.-W.} \bibnamefont{Qiu}}, \bibinfo{author}{\bibfnamefont{W.}~\bibnamefont{Vogelsang}}, \bibnamefont{and} \bibinfo{author}{\bibfnamefont{F.}~\bibnamefont{Yuan}}, \bibinfo{journal}{Phys. Rev. Lett.} \textbf{\bibinfo{volume}{97}}, \bibinfo{pages}{082002} (\bibinfo{year}{2006}), \eprint{hep-ph/0602239}.

\bibitem[{\citenamefont{Stewart et~al.}(2010{\natexlab{c}})\citenamefont{Stewart, Tackmann, and Waalewijn}}]{Stewart:2009yx}
\bibinfo{author}{\bibfnamefont{I.~W.} \bibnamefont{Stewart}}, \bibinfo{author}{\bibfnamefont{F.~J.} \bibnamefont{Tackmann}}, \bibnamefont{and} \bibinfo{author}{\bibfnamefont{W.~J.} \bibnamefont{Waalewijn}}, \bibinfo{journal}{Phys. Rev. D} \textbf{\bibinfo{volume}{81}}, \bibinfo{pages}{094035} (\bibinfo{year}{2010}{\natexlab{c}}), \eprint{0910.0467}.

\bibitem[{\citenamefont{Kang et~al.}(2012)\citenamefont{Kang, Mantry, and Qiu}}]{Kang:2012zr}
\bibinfo{author}{\bibfnamefont{Z.-B.} \bibnamefont{Kang}}, \bibinfo{author}{\bibfnamefont{S.}~\bibnamefont{Mantry}}, \bibnamefont{and} \bibinfo{author}{\bibfnamefont{J.-W.} \bibnamefont{Qiu}}, \bibinfo{journal}{Phys. Rev. D} \textbf{\bibinfo{volume}{86}}, \bibinfo{pages}{114011} (\bibinfo{year}{2012}), \eprint{1204.5469}.

\bibitem[{\citenamefont{Jouttenus et~al.}(2013)\citenamefont{Jouttenus, Stewart, Tackmann, and Waalewijn}}]{Jouttenus:2013hs}
\bibinfo{author}{\bibfnamefont{T.~T.} \bibnamefont{Jouttenus}}, \bibinfo{author}{\bibfnamefont{I.~W.} \bibnamefont{Stewart}}, \bibinfo{author}{\bibfnamefont{F.~J.} \bibnamefont{Tackmann}}, \bibnamefont{and} \bibinfo{author}{\bibfnamefont{W.~J.} \bibnamefont{Waalewijn}}, \bibinfo{journal}{Phys. Rev. D} \textbf{\bibinfo{volume}{88}}, \bibinfo{pages}{054031} (\bibinfo{year}{2013}), \eprint{1302.0846}.

\bibitem[{\citenamefont{Kang et~al.}(2013{\natexlab{a}})\citenamefont{Kang, Liu, Mantry, and Qiu}}]{Kang:2013wca}
\bibinfo{author}{\bibfnamefont{Z.-B.} \bibnamefont{Kang}}, \bibinfo{author}{\bibfnamefont{X.}~\bibnamefont{Liu}}, \bibinfo{author}{\bibfnamefont{S.}~\bibnamefont{Mantry}}, \bibnamefont{and} \bibinfo{author}{\bibfnamefont{J.-W.} \bibnamefont{Qiu}}, \bibinfo{journal}{Phys. Rev. D} \textbf{\bibinfo{volume}{88}}, \bibinfo{pages}{074020} (\bibinfo{year}{2013}{\natexlab{a}}), \eprint{1303.3063}.

\bibitem[{\citenamefont{Gaunt et~al.}(2014)\citenamefont{Gaunt, Stahlhofen, and Tackmann}}]{Gaunt:2014xga}
\bibinfo{author}{\bibfnamefont{J.~R.} \bibnamefont{Gaunt}}, \bibinfo{author}{\bibfnamefont{M.}~\bibnamefont{Stahlhofen}}, \bibnamefont{and} \bibinfo{author}{\bibfnamefont{F.~J.} \bibnamefont{Tackmann}}, \bibinfo{journal}{JHEP} \textbf{\bibinfo{volume}{04}}, \bibinfo{pages}{113} (\bibinfo{year}{2014}), \eprint{1401.5478}.

\bibitem[{\citenamefont{Bauer et~al.}(2001)\citenamefont{Bauer, Fleming, Pirjol, and Stewart}}]{Bauer:2000yr}
\bibinfo{author}{\bibfnamefont{C.~W.} \bibnamefont{Bauer}}, \bibinfo{author}{\bibfnamefont{S.}~\bibnamefont{Fleming}}, \bibinfo{author}{\bibfnamefont{D.}~\bibnamefont{Pirjol}}, \bibnamefont{and} \bibinfo{author}{\bibfnamefont{I.~W.} \bibnamefont{Stewart}}, \bibinfo{journal}{Phys. Rev. D} \textbf{\bibinfo{volume}{63}}, \bibinfo{pages}{114020} (\bibinfo{year}{2001}), \eprint{hep-ph/0011336}.

\bibitem[{\citenamefont{Bauer and Stewart}(2001)}]{Bauer:2001ct}
\bibinfo{author}{\bibfnamefont{C.~W.} \bibnamefont{Bauer}} \bibnamefont{and} \bibinfo{author}{\bibfnamefont{I.~W.} \bibnamefont{Stewart}}, \bibinfo{journal}{Phys. Lett. B} \textbf{\bibinfo{volume}{516}}, \bibinfo{pages}{134} (\bibinfo{year}{2001}), \eprint{hep-ph/0107001}.

\bibitem[{\citenamefont{Bauer et~al.}(2002{\natexlab{a}})\citenamefont{Bauer, Pirjol, and Stewart}}]{Bauer:2001yt}
\bibinfo{author}{\bibfnamefont{C.~W.} \bibnamefont{Bauer}}, \bibinfo{author}{\bibfnamefont{D.}~\bibnamefont{Pirjol}}, \bibnamefont{and} \bibinfo{author}{\bibfnamefont{I.~W.} \bibnamefont{Stewart}}, \bibinfo{journal}{Phys. Rev. D} \textbf{\bibinfo{volume}{65}}, \bibinfo{pages}{054022} (\bibinfo{year}{2002}{\natexlab{a}}), \eprint{hep-ph/0109045}.

\bibitem[{\citenamefont{Bauer et~al.}(2002{\natexlab{b}})\citenamefont{Bauer, Fleming, Pirjol, Rothstein, and Stewart}}]{Bauer:2002nz}
\bibinfo{author}{\bibfnamefont{C.~W.} \bibnamefont{Bauer}}, \bibinfo{author}{\bibfnamefont{S.}~\bibnamefont{Fleming}}, \bibinfo{author}{\bibfnamefont{D.}~\bibnamefont{Pirjol}}, \bibinfo{author}{\bibfnamefont{I.~Z.} \bibnamefont{Rothstein}}, \bibnamefont{and} \bibinfo{author}{\bibfnamefont{I.~W.} \bibnamefont{Stewart}}, \bibinfo{journal}{Phys. Rev. D} \textbf{\bibinfo{volume}{66}}, \bibinfo{pages}{014017} (\bibinfo{year}{2002}{\natexlab{b}}), \eprint{hep-ph/0202088}.

\bibitem[{\citenamefont{Beneke et~al.}(2002)\citenamefont{Beneke, Chapovsky, Diehl, and Feldmann}}]{Beneke:2002ph}
\bibinfo{author}{\bibfnamefont{M.}~\bibnamefont{Beneke}}, \bibinfo{author}{\bibfnamefont{A.~P.} \bibnamefont{Chapovsky}}, \bibinfo{author}{\bibfnamefont{M.}~\bibnamefont{Diehl}}, \bibnamefont{and} \bibinfo{author}{\bibfnamefont{T.}~\bibnamefont{Feldmann}}, \bibinfo{journal}{Nucl. Phys. B} \textbf{\bibinfo{volume}{643}}, \bibinfo{pages}{431} (\bibinfo{year}{2002}), \eprint{hep-ph/0206152}.

\bibitem[{\citenamefont{Collins et~al.}(1985)\citenamefont{Collins, Soper, and Sterman}}]{Collins:1984kg}
\bibinfo{author}{\bibfnamefont{J.~C.} \bibnamefont{Collins}}, \bibinfo{author}{\bibfnamefont{D.~E.} \bibnamefont{Soper}}, \bibnamefont{and} \bibinfo{author}{\bibfnamefont{G.~F.} \bibnamefont{Sterman}}, \bibinfo{journal}{Nucl. Phys. B} \textbf{\bibinfo{volume}{250}}, \bibinfo{pages}{199} (\bibinfo{year}{1985}).

\bibitem[{sup()}]{supplemental}
\bibinfo{note}{See Supplemental Material at [http://link.aps.org/supplemental/10.1103/rvgc-sgv7] for detailed derivations of the Sudakov factor with a zero-jettiness cut imposed, as well as additional numerical results.}

\bibitem[{\citenamefont{Ji et~al.}(2004)\citenamefont{Ji, Ma, and Yuan}}]{Ji:2004xq}
\bibinfo{author}{\bibfnamefont{X.-d.} \bibnamefont{Ji}}, \bibinfo{author}{\bibfnamefont{J.-P.} \bibnamefont{Ma}}, \bibnamefont{and} \bibinfo{author}{\bibfnamefont{F.}~\bibnamefont{Yuan}}, \bibinfo{journal}{Phys. Lett. B} \textbf{\bibinfo{volume}{597}}, \bibinfo{pages}{299} (\bibinfo{year}{2004}), \eprint{hep-ph/0405085}.

\bibitem[{\citenamefont{Qiu and Sterman}(1992)}]{Qiu:1991wg}
\bibinfo{author}{\bibfnamefont{J.-w.} \bibnamefont{Qiu}} \bibnamefont{and} \bibinfo{author}{\bibfnamefont{G.~F.} \bibnamefont{Sterman}}, \bibinfo{journal}{Nucl. Phys. B} \textbf{\bibinfo{volume}{378}}, \bibinfo{pages}{52} (\bibinfo{year}{1992}).

\bibitem[{\citenamefont{Qiu and Sterman}(1999)}]{Qiu:1998ia}
\bibinfo{author}{\bibfnamefont{J.-w.} \bibnamefont{Qiu}} \bibnamefont{and} \bibinfo{author}{\bibfnamefont{G.~F.} \bibnamefont{Sterman}}, \bibinfo{journal}{Phys. Rev. D} \textbf{\bibinfo{volume}{59}}, \bibinfo{pages}{014004} (\bibinfo{year}{1999}), \eprint{hep-ph/9806356}.

\bibitem[{\citenamefont{Rein et~al.}(2025{\natexlab{a}})\citenamefont{Rein, Schlegel, Tollk{\"u}hn, and Vogelsang}}]{Rein:2025qhe}
\bibinfo{author}{\bibfnamefont{D.}~\bibnamefont{Rein}}, \bibinfo{author}{\bibfnamefont{M.}~\bibnamefont{Schlegel}}, \bibinfo{author}{\bibfnamefont{P.}~\bibnamefont{Tollk{\"u}hn}}, \bibnamefont{and} \bibinfo{author}{\bibfnamefont{W.}~\bibnamefont{Vogelsang}}, \bibinfo{journal}{Phys. Rev. D} \textbf{\bibinfo{volume}{112}}, \bibinfo{pages}{114024} (\bibinfo{year}{2025}{\natexlab{a}}), \eprint{2503.16119}.

\bibitem[{\citenamefont{Rein et~al.}(2025{\natexlab{b}})\citenamefont{Rein, Schlegel, Tollk{\"u}hn, and Vogelsang}}]{Rein:2025pwu}
\bibinfo{author}{\bibfnamefont{D.}~\bibnamefont{Rein}}, \bibinfo{author}{\bibfnamefont{M.}~\bibnamefont{Schlegel}}, \bibinfo{author}{\bibfnamefont{P.}~\bibnamefont{Tollk{\"u}hn}}, \bibnamefont{and} \bibinfo{author}{\bibfnamefont{W.}~\bibnamefont{Vogelsang}}, \bibinfo{journal}{Phys. Rev. Lett.} \textbf{\bibinfo{volume}{135}}, \bibinfo{pages}{251901} (\bibinfo{year}{2025}{\natexlab{b}}), \eprint{2503.16097}.

\bibitem[{\citenamefont{Zhou et~al.}(2010)\citenamefont{Zhou, Yuan, and Liang}}]{Zhou:2009jm}
\bibinfo{author}{\bibfnamefont{J.}~\bibnamefont{Zhou}}, \bibinfo{author}{\bibfnamefont{F.}~\bibnamefont{Yuan}}, \bibnamefont{and} \bibinfo{author}{\bibfnamefont{Z.-T.} \bibnamefont{Liang}}, \bibinfo{journal}{Phys. Rev. D} \textbf{\bibinfo{volume}{81}}, \bibinfo{pages}{054008} (\bibinfo{year}{2010}), \eprint{0909.2238}.

\bibitem[{\citenamefont{Rein et~al.}(2023)\citenamefont{Rein, Rodini, Sch\"afer, and Vladimirov}}]{Rein:2022odl}
\bibinfo{author}{\bibfnamefont{F.}~\bibnamefont{Rein}}, \bibinfo{author}{\bibfnamefont{S.}~\bibnamefont{Rodini}}, \bibinfo{author}{\bibfnamefont{A.}~\bibnamefont{Sch\"afer}}, \bibnamefont{and} \bibinfo{author}{\bibfnamefont{A.}~\bibnamefont{Vladimirov}}, \bibinfo{journal}{JHEP} \textbf{\bibinfo{volume}{01}}, \bibinfo{pages}{116} (\bibinfo{year}{2023}), \eprint{2209.00962}.

\bibitem[{\citenamefont{Sun et~al.}(2018)\citenamefont{Sun, Isaacson, Yuan, and Yuan}}]{Sun:2014dqm}
\bibinfo{author}{\bibfnamefont{P.}~\bibnamefont{Sun}}, \bibinfo{author}{\bibfnamefont{J.}~\bibnamefont{Isaacson}}, \bibinfo{author}{\bibfnamefont{C.~P.} \bibnamefont{Yuan}}, \bibnamefont{and} \bibinfo{author}{\bibfnamefont{F.}~\bibnamefont{Yuan}}, \bibinfo{journal}{Int. J. Mod. Phys. A} \textbf{\bibinfo{volume}{33}}, \bibinfo{pages}{1841006} (\bibinfo{year}{2018}), \eprint{1406.3073}.

\bibitem[{\citenamefont{Sun and Yuan}(2013{\natexlab{a}})}]{Sun:2013dya}
\bibinfo{author}{\bibfnamefont{P.}~\bibnamefont{Sun}} \bibnamefont{and} \bibinfo{author}{\bibfnamefont{F.}~\bibnamefont{Yuan}}, \bibinfo{journal}{Phys. Rev. D} \textbf{\bibinfo{volume}{88}}, \bibinfo{pages}{034016} (\bibinfo{year}{2013}{\natexlab{a}}), \eprint{1304.5037}.

\bibitem[{\citenamefont{Echevarria et~al.}(2014)\citenamefont{Echevarria, Idilbi, Kang, and Vitev}}]{Echevarria:2014xaa}
\bibinfo{author}{\bibfnamefont{M.~G.} \bibnamefont{Echevarria}}, \bibinfo{author}{\bibfnamefont{A.}~\bibnamefont{Idilbi}}, \bibinfo{author}{\bibfnamefont{Z.-B.} \bibnamefont{Kang}}, \bibnamefont{and} \bibinfo{author}{\bibfnamefont{I.}~\bibnamefont{Vitev}}, \bibinfo{journal}{Phys. Rev. D} \textbf{\bibinfo{volume}{89}}, \bibinfo{pages}{074013} (\bibinfo{year}{2014}), \eprint{1401.5078}.

\bibitem[{\citenamefont{Bacchetta et~al.}(2022)\citenamefont{Bacchetta, Delcarro, Pisano, and Radici}}]{Bacchetta:2020gko}
\bibinfo{author}{\bibfnamefont{A.}~\bibnamefont{Bacchetta}}, \bibinfo{author}{\bibfnamefont{F.}~\bibnamefont{Delcarro}}, \bibinfo{author}{\bibfnamefont{C.}~\bibnamefont{Pisano}}, \bibnamefont{and} \bibinfo{author}{\bibfnamefont{M.}~\bibnamefont{Radici}}, \bibinfo{journal}{Phys. Lett. B} \textbf{\bibinfo{volume}{827}}, \bibinfo{pages}{136961} (\bibinfo{year}{2022}), \eprint{2004.14278}.

\bibitem[{\citenamefont{Bury et~al.}(2021)\citenamefont{Bury, Prokudin, and Vladimirov}}]{Bury:2021sue}
\bibinfo{author}{\bibfnamefont{M.}~\bibnamefont{Bury}}, \bibinfo{author}{\bibfnamefont{A.}~\bibnamefont{Prokudin}}, \bibnamefont{and} \bibinfo{author}{\bibfnamefont{A.}~\bibnamefont{Vladimirov}}, \bibinfo{journal}{JHEP} \textbf{\bibinfo{volume}{05}}, \bibinfo{pages}{151} (\bibinfo{year}{2021}), \eprint{2103.03270}.

\bibitem[{\citenamefont{Kang and Qiu}(2009{\natexlab{b}})}]{Kang:2008ey}
\bibinfo{author}{\bibfnamefont{Z.-B.} \bibnamefont{Kang}} \bibnamefont{and} \bibinfo{author}{\bibfnamefont{J.-W.} \bibnamefont{Qiu}}, \bibinfo{journal}{Phys. Rev. D} \textbf{\bibinfo{volume}{79}}, \bibinfo{pages}{016003} (\bibinfo{year}{2009}{\natexlab{b}}), \eprint{0811.3101}.

\bibitem[{\citenamefont{Zhou et~al.}(2009)\citenamefont{Zhou, Yuan, and Liang}}]{Zhou:2008mz}
\bibinfo{author}{\bibfnamefont{J.}~\bibnamefont{Zhou}}, \bibinfo{author}{\bibfnamefont{F.}~\bibnamefont{Yuan}}, \bibnamefont{and} \bibinfo{author}{\bibfnamefont{Z.-T.} \bibnamefont{Liang}}, \bibinfo{journal}{Phys. Rev. D} \textbf{\bibinfo{volume}{79}}, \bibinfo{pages}{114022} (\bibinfo{year}{2009}), \eprint{0812.4484}.

\bibitem[{\citenamefont{Vogelsang and Yuan}(2009)}]{Vogelsang:2009pj}
\bibinfo{author}{\bibfnamefont{W.}~\bibnamefont{Vogelsang}} \bibnamefont{and} \bibinfo{author}{\bibfnamefont{F.}~\bibnamefont{Yuan}}, \bibinfo{journal}{Phys. Rev. D} \textbf{\bibinfo{volume}{79}}, \bibinfo{pages}{094010} (\bibinfo{year}{2009}), \eprint{0904.0410}.

\bibitem[{\citenamefont{Braun et~al.}(2009)\citenamefont{Braun, Manashov, and Pirnay}}]{Braun:2009mi}
\bibinfo{author}{\bibfnamefont{V.~M.} \bibnamefont{Braun}}, \bibinfo{author}{\bibfnamefont{A.~N.} \bibnamefont{Manashov}}, \bibnamefont{and} \bibinfo{author}{\bibfnamefont{B.}~\bibnamefont{Pirnay}}, \bibinfo{journal}{Phys. Rev. D} \textbf{\bibinfo{volume}{80}}, \bibinfo{pages}{114002} (\bibinfo{year}{2009}), \bibinfo{note}{[Erratum: Phys.Rev.D 86, 119902 (2012)]}, \eprint{0909.3410}.

\bibitem[{\citenamefont{Ma and Sang}(2011)}]{Ma:2011nd}
\bibinfo{author}{\bibfnamefont{J.~P.} \bibnamefont{Ma}} \bibnamefont{and} \bibinfo{author}{\bibfnamefont{H.~Z.} \bibnamefont{Sang}}, \bibinfo{journal}{JHEP} \textbf{\bibinfo{volume}{04}}, \bibinfo{pages}{062} (\bibinfo{year}{2011}), \eprint{1102.2679}.

\bibitem[{\citenamefont{Schafer and Zhou}(2012)}]{Schafer:2012ra}
\bibinfo{author}{\bibfnamefont{A.}~\bibnamefont{Schafer}} \bibnamefont{and} \bibinfo{author}{\bibfnamefont{J.}~\bibnamefont{Zhou}}, \bibinfo{journal}{Phys. Rev. D} \textbf{\bibinfo{volume}{85}}, \bibinfo{pages}{117501} (\bibinfo{year}{2012}), \eprint{1203.5293}.

\bibitem[{\citenamefont{Ma and Wang}(2012)}]{Ma:2012xn}
\bibinfo{author}{\bibfnamefont{J.~P.} \bibnamefont{Ma}} \bibnamefont{and} \bibinfo{author}{\bibfnamefont{Q.}~\bibnamefont{Wang}}, \bibinfo{journal}{Phys. Lett. B} \textbf{\bibinfo{volume}{715}}, \bibinfo{pages}{157} (\bibinfo{year}{2012}), \eprint{1205.0611}.

\bibitem[{\citenamefont{Kang and Qiu}(2012)}]{Kang:2012em}
\bibinfo{author}{\bibfnamefont{Z.-B.} \bibnamefont{Kang}} \bibnamefont{and} \bibinfo{author}{\bibfnamefont{J.-W.} \bibnamefont{Qiu}}, \bibinfo{journal}{Phys. Lett. B} \textbf{\bibinfo{volume}{713}}, \bibinfo{pages}{273} (\bibinfo{year}{2012}), \eprint{1205.1019}.

\bibitem[{\citenamefont{Sun and Yuan}(2013{\natexlab{b}})}]{Sun:2013hua}
\bibinfo{author}{\bibfnamefont{P.}~\bibnamefont{Sun}} \bibnamefont{and} \bibinfo{author}{\bibfnamefont{F.}~\bibnamefont{Yuan}}, \bibinfo{journal}{Phys. Rev. D} \textbf{\bibinfo{volume}{88}}, \bibinfo{pages}{114012} (\bibinfo{year}{2013}{\natexlab{b}}), \eprint{1308.5003}.

\bibitem[{\citenamefont{Zhou}(2015)}]{Zhou:2015lxa}
\bibinfo{author}{\bibfnamefont{J.}~\bibnamefont{Zhou}}, \bibinfo{journal}{Phys. Rev. D} \textbf{\bibinfo{volume}{92}}, \bibinfo{pages}{074016} (\bibinfo{year}{2015}), \eprint{1507.02819}.

\bibitem[{\citenamefont{del Rio et~al.}(2024)\citenamefont{del Rio, Prokudin, Scimemi, and Vladimirov}}]{delRio:2024vvq}
\bibinfo{author}{\bibfnamefont{O.}~\bibnamefont{del Rio}}, \bibinfo{author}{\bibfnamefont{A.}~\bibnamefont{Prokudin}}, \bibinfo{author}{\bibfnamefont{I.}~\bibnamefont{Scimemi}}, \bibnamefont{and} \bibinfo{author}{\bibfnamefont{A.}~\bibnamefont{Vladimirov}}, \bibinfo{journal}{Phys. Rev. D} \textbf{\bibinfo{volume}{110}}, \bibinfo{pages}{016003} (\bibinfo{year}{2024}), \eprint{2402.01836}.

\bibitem[{foo()}]{footnote}
\bibinfo{note}{The total integrated luminosity $\mathcal{L} = 780 \,\text{pb}^{-1}$ combines data from the 2017 RHIC run ($350 \, \text{pb}^{-1}$ with polarization $P \approx 55\%$) and the 2022 run ($430 \, \text{pb}^{-1}$ with $P \approx 50\%$). An effective polarization of $P = 53\%$ is used in our calculation.}

\bibitem[{\citenamefont{Kang et~al.}(2013{\natexlab{b}})\citenamefont{Kang, Lee, and Stewart}}]{Kang:2013nha}
\bibinfo{author}{\bibfnamefont{D.}~\bibnamefont{Kang}}, \bibinfo{author}{\bibfnamefont{C.}~\bibnamefont{Lee}}, \bibnamefont{and} \bibinfo{author}{\bibfnamefont{I.~W.} \bibnamefont{Stewart}}, \bibinfo{journal}{Phys. Rev. D} \textbf{\bibinfo{volume}{88}}, \bibinfo{pages}{054004} (\bibinfo{year}{2013}{\natexlab{b}}), \eprint{1303.6952}.

\bibitem[{\citenamefont{Nadel-Turonski}(2025)}]{Nadel-Turonski:2025sfv}
\bibinfo{author}{\bibfnamefont{P.}~\bibnamefont{Nadel-Turonski}}, \bibinfo{journal}{Acta Phys. Polon. Supp.} \textbf{\bibinfo{volume}{18}}, \bibinfo{pages}{1} (\bibinfo{year}{2025}).

\bibitem[{\citenamefont{Bijl et~al.}(2024)\citenamefont{Bijl, Niedenzu, and Waalewijn}}]{Bijl:2023dux}
\bibinfo{author}{\bibfnamefont{P.}~\bibnamefont{Bijl}}, \bibinfo{author}{\bibfnamefont{S.}~\bibnamefont{Niedenzu}}, \bibnamefont{and} \bibinfo{author}{\bibfnamefont{W.~J.} \bibnamefont{Waalewijn}}, \bibinfo{journal}{Phys. Rev. D} \textbf{\bibinfo{volume}{109}}, \bibinfo{pages}{014011} (\bibinfo{year}{2024}), \eprint{2307.02521}.

\bibitem[{\citenamefont{Collins and Rogers}(2015)}]{Collins:2014jpa}
\bibinfo{author}{\bibfnamefont{J.}~\bibnamefont{Collins}} \bibnamefont{and} \bibinfo{author}{\bibfnamefont{T.}~\bibnamefont{Rogers}}, \bibinfo{journal}{Phys. Rev. D} \textbf{\bibinfo{volume}{91}}, \bibinfo{pages}{074020} (\bibinfo{year}{2015}), \eprint{1412.3820}.

\end{thebibliography}

\clearpage
\appendix
\onecolumngrid
\setcounter{equation}{0}
\renewcommand{\theequation}{S-\arabic{equation}}
 \allowdisplaybreaks
\section*{Supplemental Material} 

\subsection*{Perturbative QCD phase space analysis}

We provide a derivation of the joint NLL Sudakov factor $S_P(b)$ using a direct analysis of emission phase space within perturbative QCD. We begin with the standard NLL expression for the unpolarized Drell-Yan cross section in the CSS formalism
\begin{align} \label{eq:Supp_CSS_baseline}
 \frac{\d \sigma_{UU}}{ \d y \, \d^2 \vec{q}_\perp} = \sigma_0 \int_0^\infty\frac{b \, \d b}{2\pi} J_0 (b\,q_\perp) \sum\limits_{q,q'} |V_{qq'}|^2 f_{q}(x_q,\mu_b)f_{q'}(x_{q'},\mu_b)e^{-S_{\rm P}^{\rm TMD}(b)}\,,
\end{align}
where $\sigma_0$ is the Born cross section, $f_i$ are collinear PDFs evaluated at $\mu_b = 2e^{-\gamma_E}/b$, and $S_{\rm P}^{\rm TMD}(b)$ is the standard TMD Sudakov factor arising from resumming logarithms of $Q/q_\perp$. This standard factor originates from exponentiating single gluon emissions.

To understand the structure of $S_{\rm P}^{\rm TMD}(b)$, consider the one-loop real emission correction contributing logarithms. For clarity, we consider gluon radiation in hemisphere `a' (with respect to the beam axis) which implies $l^- < l^+$ where $l^- \equiv n_{a}\cdot l /\sqrt{2}$ and $l^+\equiv n_{b}\cdot l/\sqrt{2}$.  This configuration imposes the kinematic constraint $l_\perp/\sqrt{2}<l^+<p_q^+=Q/\sqrt{2}$  with $l_\perp=|\vec l_\perp|$, due to the relation $l^+=l_\perp e^{y_g}/\sqrt{2}$ and $y_g>0$ in the hemisphere `a'. The real-emission correction includes collinear and soft contributions, governed by the unregularized DGLAP splitting kernel, $\int_0^{1- l_\perp \! /Q} \! \d z\,  (1+z^2)/(1-z)$, with $1-z=l^+/p_q^+$. Notably, this kinematic constraint provides a natural regularization of the light-cone divergence as $z \rightarrow 1$. The full real emission correction can be decomposed as
\begin{eqnarray} 
\tilde S_{\rm P}^{\rm real}=\frac{\alpha_s C_F}{2\pi^2} \!\! \int_0^{Q^2} \! \frac{\d^2  \vec l_\perp}{ l^{2}_\perp}\!   \int_0^{1} \d z \left\{ \left [
  \frac{1+z^2}{(1-z)_+}+\frac{3}{2}\delta(1-z)  \right ]  -\delta(1-z)\left [ \frac{3}{2}+\int_0^{1- l_\perp/Q }\d z' \,\frac{2}{1-z'}\right ] \right \}\,,
  \label{decom}
\end{eqnarray}
where the first integral in the brace drives the DGLAP evolution of PDFs, while the rest terms generate the leading double and single logarithms characteristic of TMD factorization. Including the virtual correction in the impact parameter $b$ space gives, 
\begin{eqnarray} 
\frac{\alpha_s C_F}{2 \pi^2} 
\int_0^Q \frac{d^2 l_{\perp}}{l_{\perp}^2}
\left(\ln \frac{Q^2}{l_{\perp}^2} - \frac{3}{2}\right)
\left[e^{i l_{\perp} \cdot b} - 1\right]
&\approx&
\frac{\alpha_s C_F}{2 \pi}
\left(
-\frac{1}{2} \ln^2 \frac{Q^2}{\mu_b^2}
+ \frac{3}{2} \ln \frac{Q^2}{\mu_b^2}
\right)\,.
\end{eqnarray}
According to the standard Collins-Soper-Sterman (CSS) formalism~\cite{Collins:2011zzd}, these logarithmic contributions exponentiate into the perturbative Sudakov factor in impact parameter $b$ space:
\begin{eqnarray} 
S_{\text P}^{\rm TMD}(b)\!=\!\frac{C_F}{\pi} \int_{\mu_b^2}^{Q^2} \!\!
 \frac{\d \mu^2}{\mu^2} \left(
 \ln \frac{ Q^2}{\mu^2}\!-\! \frac{3}{2}\right )  \alpha_s(\mu)\,. 
\end{eqnarray}

Imposing a veto with $\tau_0 \ll 1$ restricts the phase space available for initial-state radiation, significantly modifying the logarithmic structure of the cross section. We separately compute the different logarithmic terms of the soft factor at one-loop order arising from different phase-space regions,
\begin{eqnarray} 
S_{\text P}^{\rm 1}\!=\! S_{\text P}^{\rm 1a}+S_{\text P}^{\rm 1b}+S_{\text P}^{\rm1c}+S_{\text P}^{\rm 1d}+ S_{\text P}^{\rm 1e}\,. 
\end{eqnarray}
First of all, the veto  modifies the minimal value of $l^+$ to $l^2_\perp  Q/(\sqrt{2}t)$, where $t=\tau_0 Q^2$, and bounds the integration of radiated gluon transverse momentum by $t$. Under these constraints, the leading double-logarithmic contribution from gluon radiation in hemisphere `a' takes the following modified form
\begin{eqnarray} 
S_{\text P}^{\rm 1a}=\frac{\alpha_sC_F}{\pi^2}\!\! \int_{0}^{ t}
\! \frac{\d^2 \vec l_\perp}{l^2_\perp}\!\int^{p_q^+}_{\frac{l^2_\perp  Q}{\sqrt{2}t }} \frac{\d l^+}{l^+} \left (e^{i \vec b \cdot \vec l_\perp} -1 \right )\approx -\frac{\alpha_sC_F}{2\pi}   \ln^2 \!\frac{t}{\mu_b^2}\,.  \label{double}
\end{eqnarray}
However, when $l^+ < l_\perp/\sqrt{2}$, the gluon is emitted into the opposite hemisphere. Requiring consistency with $l_\perp > l^2_\perp Q/t$ leads to the condition $l^2_\perp < \tau_0 t$. To avoid double-counting in this phase space, we subtract the following contribution
\begin{eqnarray} 
 S_{\text P}^{\rm 1b}=-\frac{\alpha_s C_F}{\pi^2} \!\! \int_{0}^{\tau_0t}\!\!
\frac{\d^2 \vec l_\perp}{l^2_\perp} 
  \!\! \int^{\frac{l_\perp}{\sqrt{2}}}_{\frac{l^2_\perp  Q}{\sqrt{2}t }} \frac{\d l^+}{l^+} \left (e^{i \vec b \cdot \vec l_\perp} -1 \right ) \approx \frac{\alpha_sC_F}{4\pi} \ln^2 \!\frac{\tau_0 t}{\mu_b^2}\,.
\end{eqnarray}
Following the procedure outlined in Eq.~\eqref{decom}, once the TMD is matched onto the PDF at its natural scale $t$, a single logarithm remains to be resummed,
\begin{eqnarray} 
S_{\text P}^{\rm 1c}=\frac{\alpha_sC_F}{\pi^2}\frac{3}{2} \int_{0}^{ t}
\! \frac{\d^2 \vec l_\perp}{l^2_\perp}\left (e^{i\vec b \cdot \vec l_\perp} -1 \right )\approx -\frac{\alpha_sC_F}{\pi} \frac{3}{2}  \ln \frac{t}{\mu_b^2}\,.
\end{eqnarray}
The virtual corrections are unaffected by the 0-jettiness. In addition to the components paired with the real emission, there are further contributions from the two phase space regions. In the region $\tau_0 t < l^2_\perp < t$ with $l_\perp/\sqrt{2} < l^+ < l^2_\perp Q/(\sqrt{2}t)$, the real emissions are vetoed while the virtual corrections remain
\begin{eqnarray} 
S_{\text P}^{\rm 1d} =\frac{\alpha_s C_F}{\pi^2}  \int_{\tau_0t }^{t}
\frac{\d^2 \vec l_\perp}{l^2_\perp}\int_{\frac{l_\perp}{\sqrt{2}}}^{\frac{l^2_\perp  Q}{\sqrt{2}t }} \frac{\d l^+}{l^+} = \frac{\alpha_s C_F}{4\pi} \ln^2 \frac{1}{\tau_0}\,.
\end{eqnarray}
The virtual contribution above the scale $t$ to single-logarithmic accuracy is given by
\begin{eqnarray} 
 S_{\text P}^{\rm 1e}=-\frac{\alpha_s C_F}{2\pi} \left [\frac{1}{2 }\ln^2 \frac{Q^2}{t} - \frac{3}{2} \ln \frac{Q^2}{t} \right ]\,,
\end{eqnarray}
yielding a global suppression factor after carrying out all-order resummation.

Combining the contributions from two hemispheres and exponentiating the one-loop result $S_{\text P}^{\rm 1}$, one eventually ends up with the following Sudakov factor for the isolated Drell-Yan-like process, 
\begin{eqnarray} \label{eq:Sud}
&&\!\!\!\!\!\!\!\!\!\!\!\!
S_{\text P}(b)\!=\!\frac{C_F}{\pi}\! \left [\int_{\mu_b^2}^{t} \!\!
 \frac{\d \mu^2}{\mu^2}  
 \left ( 2\ln \frac{t}{\mu^2} -\frac{3}{2}\right ) -\int_{\mu_b^2}^{\tau_0t} \!\!
 \frac{\d \mu^2}{\mu^2} 
 \ln \frac{\tau_0 t}{\mu^2}  + \int_{t}^{Q^2} \!\!
 \frac{\d \mu^2}{\mu^2} \left(
 \ln \frac{ Q^2}{\mu^2}\!-\! \frac{3}{2}\right ) +\!\int^{t}_{\tau_0t} 
 \!\! \frac{\d \mu^2}{\mu^2} 
 \ln \frac{\mu^2}{ \tau_0t}   \right ]\alpha_s(\mu)\,.
\end{eqnarray}

\subsection*{Formulation in SCET}

In this supplemental material, we present a rigorous derivation of the Sudakov factor within the framework of Soft-Collinear Effective Theory (SCET). While the derivation in the main text offers an intuitive and computationally efficient approach, it lacks the systematic power to resum sub-leading logarithmic terms. The SCET framework provides a more robust theoretical foundation for such resummation. Previous work on joint $0$-jettiness and TMD resummation within SCET~\cite{Procura:2014cba, Lustermans:2019plv} serves as an excellent foundation for our analysis. Here, we elucidate the connection between these established results and the resummation formula presented in the main text.

\begin{figure}[!thb]
	\centering
	\includegraphics[scale=0.5]{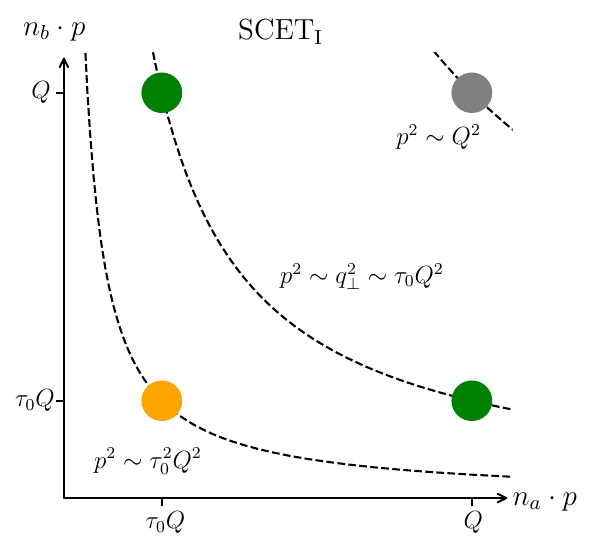}~
	\includegraphics[scale=0.5]{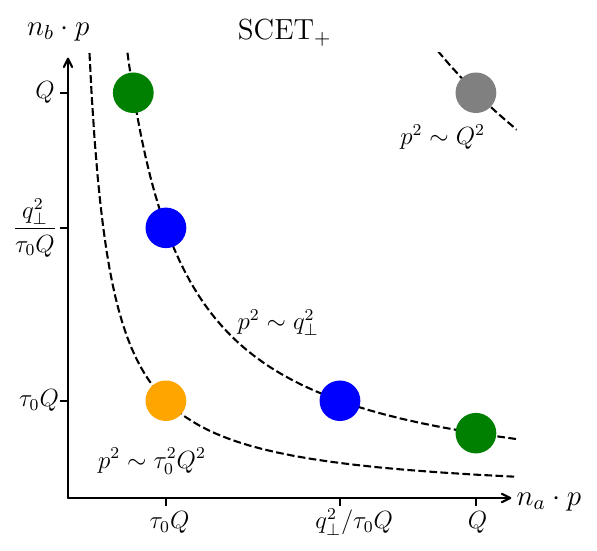}~
	\includegraphics[scale=0.5]{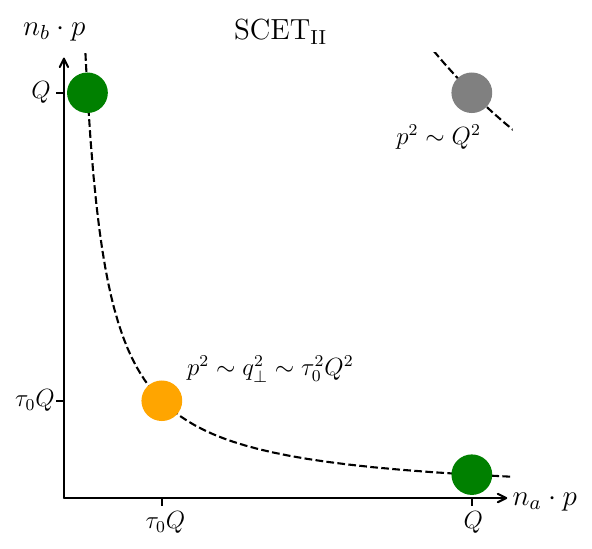}~
	\caption{The kinematic modes in three regimes for SCET$_{\rm I}$, SCET$_{\rm +}$, and SCET$_{\rm II}$, respectively. The hard (gray), collinear (green), soft (orange), and collinear-soft (blue) modes are indicated. The light-cone momentum components and the invariant masses are shown.}
	\label{fig:SCET}
\end{figure}

The joint 0-jettiness and TMD factorization formulas within SCET are organized according to three distinct scale hierarchies:
\begin{itemize}
	\item SCET$_{\rm I}$: $\tau_0^2 Q^2 \ll q_\perp^2 \sim \tau_0 Q^2 \ll Q^2$\,;
	\item SCET$_{\rm +}$: $\tau_0^2 Q^2 \ll q_\perp^2 \ll \tau_0 Q^2 \ll Q^2$\,;
	\item SCET$_{\rm II}$: $\tau_0^2 Q^2 \sim q_\perp^2 \ll \tau_0 Q^2 \ll Q^2$\,;
\end{itemize}
These regimes correspond to different momentum scalings of the hard, collinear, soft and collinear-soft modes. Fig.~\ref{fig:SCET} provides a schematic summary of the mode structure in each regime.

\subsubsection*{SCET$_{\rm I}$ Regime}


In the SCET$_{\rm I}$ regime, defined by the hierarchy $\tau_0^2 Q^2\ll q_\perp^2 \sim \tau_0 Q^2\ll Q^2$, the factorization formula for the differential cross section is given by:
\begin{align} \label{eq:factorization_scet1}
\frac{\d \sigma_{\rm I}(\tau_0)}{ \d y\, \d^2 \vec q_\perp}
&= \sigma_0\, H(Q, \mu)  \sum_{q,q'} C_{qq'}
\int_0^{\tau_0 Q}\! \d \Tau
 \int\! \d t_q\int\! \d^2 \vec k_q \, {\mathcal B}_{q/p}(t_q, x_q, \vec k_q,  \mu)
 \int\! \d t_{q'}\int\! \d^2 \vec k_{q'} \, {\mathcal B}_{{q'}/p}(t_{q'}, x_{q'}, \vec k_{q'},  \mu)
 \nn\\ &\quad \times
 \int\! \d k\,{\mathcal S}(k,  \mu) \,
\delta^{(2)} \left( {\vec q}_\perp - \vec k_{q} - \vec k_{q'} \right)
\delta\left( \Tau - \frac{t_q}{Q} - \frac{t_{q'}}{Q}-k\right)\nn\\
&= \sigma_0\, H(Q, \mu)  \sum_{q,q'} C_{qq'}
\int_0^{\tau_0 Q}\! \d \Tau \int\! \frac{ e^{s \Tau} \d s}{2\pi i}
 \int_0^\infty\! \frac{b\, \d b  }{2\pi} J_0(b\,q_\perp) \, {\mathcal B}_{q/p}(s/Q, x_q, b,  \mu)
 \, {\mathcal B}_{q'/p}(s/Q, x_{q'}, b,  \mu)\,
 {\mathcal S}(s,  \mu) 
\,,\end{align}
where $\sigma_0=\sqrt{2}\pi G_F Q^2 /(sN_c)$, and $C_{qq'}=|V_{qq'}|^2$ for $W^{\pm}$ production and $C_{qq'}=(V_{q}^2+A_{q}^2)\delta_{q{\bar q}'}$ for $Z^{0}$ production. Here, $H(Q, \mu)$ is the hard function, encoding the short-distance partonic scattering at the hard scale $Q$. The function ${\mathcal B}_{q/p}(s/Q, x_q, b, \mu)$ denotes the double-differential beam function for parton $q$ originating from proton $p$; it incorporates the effects of initial-state radiation and is expressed in $b$-space (Fourier conjugate to the transverse momentum $\vec{k}_q$) and $s$-space (Laplace conjugate to the 0-jettiness variable $t_q/Q$). ${\mathcal S}( s, \mu)$ is the soft function, which accounts for soft gluon emissions contributing to the $0$-jettiness measurement. The factorization formula is first presented in momentum space, and then in $b$- and $s$-space, where the convolutions over transverse momenta and $0$-jettiness are replaced by simple products. The Bessel function $J_0(b\,q_\perp)$ arises from the Fourier transform between $b$-space and transverse momentum space.

The evolution of these functions with the renormalization scale $\mu$ is governed by the following renormalization group (RG) equations in $b$-space and $s$-space:
\begin{align}
    \mu \frac{\d}{\d\mu} {\mathcal B}_{q/p}(s/Q, x_q, b,\mu)&=\gamma_{\mathcal B}^q(s/Q,\mu) {\mathcal B}_{q/p}(s/Q, x_q, b,\mu) \,, \nn\\
    \gamma_{\mathcal B}^q(s/Q,\mu)&=-2\Gamma_{\rm cusp}^q(\alpha_s) \ln\frac{ Q}{\mu^2 s e^{\gamma_E}}+\gamma_{\mathcal B}^q(\alpha_s)\,,\\
    \mu \frac{\d}{\d\mu} {\mathcal S}(s, \mu) &=\gamma_{\mathcal S}^q( s,\mu) {\mathcal S}(s, \mu)  \,, \nn\\
    \gamma_{\mathcal S}^q( s,\mu)&=2\Gamma_{\rm cusp}^q(\alpha_s) \ln\frac{ 1}{\mu^2 s^2 e^{2\gamma_E}}+\gamma_{\mathcal S}^q(\alpha_s)\,,\\
      \mu \frac{\d}{\d\mu} H(Q, \mu) &=\gamma_H^q(Q,\mu) H(Q, \mu)  \,, \nn\\
    \gamma_H^q(Q,\mu)&=2\Gamma_{\rm cusp}^q(\alpha_s) \ln\frac{ Q^2}{\mu^2}+2\gamma_V^q(\alpha_s)\,.
\end{align}
These equations describe how the double-differential beam functions (${\mathcal B}_{q/p}$), the beam thrust soft function (${\mathcal S}$), and the hard function ($H$) change with the renormalization scale $\mu$, as dictated by their respective anomalous dimensions $\gamma_{\mathcal B}^q$, $\gamma_{\mathcal S}^q$, and $2\gamma_V^q$. The cusp anomalous dimension $\Gamma_{\rm cusp}^q$ and non-cusp anomalous dimensions $\gamma_{\mathcal B}^q(\alpha_s)$, $\gamma_{\mathcal S}^q(\alpha_s)$, $2\gamma_V^q(\alpha_s)$ appear in these expressions.
All the corresponding anomalous dimensions in this section are shown in the following section. The canonical choices for the renormalization scales associated with each function in this regime are set to minimize large logarithms in perturbative calculations:
\begin{align} \label{eq:canonical_scales_scet1}
\mu_H^{\rm I} \sim Q
\,, \qquad
\mu_{\mathcal B}^{\rm I} \sim \sqrt{\tau_0} Q
\,, \qquad
\mu_{\mathcal S}^{\rm I} \sim \tau_0 Q 
\,.\end{align}
These scales reflect the characteristic momentum scales of the hard, collinear (beam), and soft interactions, respectively. After implementing the one-loop
 anomalous dimensions and integrating $s$ and $\tau$, we define the perturbative Sudakov factor $S_{\text P}(b)$ as
\begin{align}
S_{\text P}(b)\!&=\!\frac{C_F}{\pi}\! 
\int_0^{\tau_0 Q}\! \d \Tau \int\! \frac{ e^{s \Tau} \d s}{2\pi i}\left [
 \int_{\tau_0 Q^2}^{\mu_b^2} \!\!
  \frac{\d \mu^2}{\mu^2} 
 \left(2\ln \frac{ Q}{\mu^2 s e^{\gamma_E}}-\frac{3}{2}\right) -
 \int_{\tau_0^2 Q^2}^{\mu_b^2} \!\!
  \frac{\d \mu^2}{\mu^2} 
 \ln \frac{ 1}{\mu^2 s^2 e^{2\gamma_E}}  
 +\! \int_{\mu_b^2}^{Q^2} \!\!
 \frac{\d \mu^2}{\mu^2} \left(
 \ln \frac{ Q^2}{\mu^2}\!-\! \frac{3}{2}\right )   \right ]\!\!  \alpha_s(\mu)  \nn\\
\!&=\!\frac{C_F}{\pi}\! \left [
 \int_{\tau_0 Q^2}^{\mu_b^2} \!\!
  \frac{\d \mu^2}{\mu^2} 
 \left(2\ln \frac{\tau_0 Q^2}{\mu^2}-\frac{3}{2}\right) -
 \int_{\tau_0^2 Q^2}^{\mu_b^2} \!\!
  \frac{\d \mu^2}{\mu^2} 
 \ln \frac{\tau_0^2 Q^2}{\mu^2 }  
 +\! \int_{\mu_b^2}^{Q^2} \!\!
 \frac{\d \mu^2}{\mu^2} \left(
 \ln \frac{ Q^2}{\mu^2}\!-\! \frac{3}{2}\right )   \right ]\!\!  \alpha_s(\mu) \,.    
\end{align}
This factor exponentiates the dominant logarithmic contributions arising from the evolution of the functions from their canonical scales (Eq.~\eqref{eq:canonical_scales_scet1}) to a common scale $\mu_b = b_0/b$ with $b_0\equiv 2e^{-\gamma_{\rm E}}$.

\subsubsection*{SCET$_{\rm II}$ Regime}

Next, we consider the SCET$_{\rm II}$ regime, characterized by the scale hierarchy $\tau_0^2 Q^2\sim q_\perp^2 \ll \tau_0 Q^2 \ll Q^2$. In this regime, the factorization formula takes the form:
\begin{align} \label{eq:factorization_scet2}
\frac{\d \sigma_{\rm II}(\tau_0)}{ \d y\, \d^2 \vec q_\perp}
&= \sigma_0\, H(Q, \mu)  \sum_{q,q'} C_{qq'}
\int_0^{\tau_0 Q}\! \d \Tau
 \int\! \d^2 \vec k_q \, {B}_{q/p}( x_q, \vec k_q, \mu,\nu/\omega_q)
 \int\! \d^2 \vec k_{q'} \, {B}_{q'/p}( x_{q'}, \vec k_{q'}, \mu,\nu/\omega_{q'})
 \\ &\quad \times
\int\! \d k\, S( k, \vec k_s , \mu, \nu) \,
\delta^{(2)}\left( {\vec q}_\perp - \vec k_{q} - \vec k_{q'} -\vec k_s \right)
\delta\left( \Tau -k \right) \nn \\
&= \sigma_0\, H(Q, \mu)  \sum_{q,q'} C_{qq'}
\int_0^{\tau_0 Q}\! \d \Tau \int\! \frac{ e^{s \Tau} \d s}{2\pi i}
  \int_0^\infty \! \frac{b\, \d b  }{2\pi} J_0(b\,q_\perp) \, {B}_{q/p}( x_q, b, \mu,\nu/\omega_q)
 \, {B}_{q'/p}( x_{q'}, b, \mu,\nu/\omega_{q'})\,
 S(s, b, \mu,\nu) 
\,, \nn\end{align}
with $\omega_q=Qe^{+y}$ and $\omega_{q'}=Qe^{-y}$.
In this expression, $H(Q, \mu)$ is the hard function. ${B}_{q/p}(x_q, b, \mu,\nu/\omega_q)$ represents the standard TMD PDF, which depends on both the renormalization scale $\mu$ and the rapidity scale $\nu$. $S(s, b,  \mu,\nu)$ is the double-differential soft function, also depending on $\mu$ and $\nu$. 

The RG and rapidity-RG (RRG) equations for the functions in this regime, describing their evolution with respect to both $\mu$ and $\nu$, are:
\begin{align}
    \mu \frac{\d}{\d\mu} {B}_{q/p}(x_q, b, \mu,\nu/\omega_q)&={\gamma}_B^q(\mu,\nu/\omega_q) {B}_{q/p}( x_q, b, \mu,\nu/\omega_q) \,, \nn\\
    \nu \frac{\d}{\d\nu} {B}_{q/p}( x_q, b, \mu,\nu/\omega_q)&=-\frac{1}{2}{\gamma}_\nu^q(b,\mu) {B}_{q/p}( x_q, b, \mu,\nu/\omega_q) \,, \nn\\
    \gamma_B^q(\mu,\nu/\omega_q)&=\Gamma_{\rm cusp}^q(\alpha_s) \ln\frac{\nu^2}{\omega_q^2}+\gamma_B^q(\alpha_s)\,,\nn\\
    {\gamma}_\nu^q(b,\mu)&=2\left[\int_{\mu^2}^{\mu_b^2}\frac{\d\mu'^2}{\mu'^2}\Gamma_{\rm cusp}^q(\alpha_s)+\gamma_{ r}^q(\alpha_s)\right]\,,\\
    \mu \frac{\d}{\d\mu} S( s,b, \mu,\nu) &=\gamma_S^q(\mu,\nu) S( s,b, \mu,\nu)  \,, \nn\\
    \nu \frac{\d}{\d\nu} S( s,b, \mu,\nu) &={\gamma}_\nu^q(b,\mu) S( s,b, \mu,\nu)  \,, \nn\\
    \gamma_S^q(\mu,\nu)&=2\Gamma_{\rm cusp}^q(\alpha_s) \ln\frac{\mu^2}{\nu^2}+\gamma_S^q(\alpha_s)\,,\\
      \mu \frac{\d}{\d\mu} H(Q, \mu) &=\gamma_H^q(Q,\mu) H(Q, \mu)  \,, \nn\\
    \gamma_H^q(Q,\mu)&=2\Gamma_{\rm cusp}^q(\alpha_s) \ln\frac{ Q^2}{\mu^2}+2\gamma_V^q(\alpha_s)\,.
\end{align}
These equations detail the scale dependence of the TMD beam functions ($B_{q/p}$), the double-differential soft function ($S$), and the hard function ($H$). The TMD functions evolve with both the renormalization scale $\mu$ (governed by anomalous dimensions $\gamma_B^q, \gamma_S^q$) and the rapidity scale $\nu$ (governed by the rapidity anomalous dimension $\gamma_\nu^q$). 
The canonical scales for the functions in the SCET$_{\rm II}$ regime are:
\begin{align} \label{eq:canonical_scales_scet2}
\mu_H^{\rm II} \sim Q
\,, \qquad
&\mu_B^{\rm II} \sim \mu_b
\,, \qquad
\mu_S^{\rm II} \sim \mu_b\,, \nn\\
&\nu_B^{\rm II} \sim Q
\,, \qquad\ 
\nu_S^{\rm II} \sim \mu_b
\,.\end{align}
Here, $\mu_b$ is the scale associated with the transverse momentum $q_\perp$.

Therefore, the perturbative Sudakov factor for the SCET$_{\rm II}$ regime, which resums logarithms associated with the evolution from canonical scales, is:
\begin{align}
S_{\text P}(b)\!&=\!\frac{C_F}{\pi}\!
\int_0^{\tau_0 Q}\! \d \Tau \int\! \frac{ e^{s \Tau} \d s}{2\pi i}\left [
 \! \int_{\mu_b^2}^{Q^2} \!\!
 \frac{\d \mu^2}{\mu^2} \left(
 \ln \frac{ Q^2}{\mu^2}\!-\! \frac{3}{2}\right )   \right ]\!\!  \alpha_s(\mu)\nn\\
\!&=\!\frac{C_F}{\pi}\! \left [
 \! \int_{\mu_b^2}^{Q^2} \!\!
 \frac{\d \mu^2}{\mu^2} \left(
 \ln \frac{ Q^2}{\mu^2}\!-\! \frac{3}{2}\right )   \right ]\!\!  \alpha_s(\mu) \,.   
\end{align}
This factor primarily accounts for the evolution of the hard function and the $\mu$-evolution part of the TMD functions from $\mu_b$ to $Q$. This regime corresponds to standard TMD evolution.

\subsubsection*{SCET$_{\rm +}$ Regime}

The third regime, SCET$_{\rm +}$, is defined by the hierarchy $\tau_0^2 Q^2\ll q_\perp^2 \ll \tau_0 Q^2$. The factorization formula in this intermediate regime is:
\begin{align} \label{eq:factorization_scet+}
\frac{\d \sigma_{\rm +}(\tau_0)}{ \d y\, \d^2 \vec q_\perp}
&= \sigma_0\, H(Q, \mu)  \sum_{q,q'} C_{qq'}
\int_0^{\tau_0 Q}\! \d \Tau
 \int\! \d^2 \vec k_q \, {B}_{q/p}( x_q, \vec k_q, \mu,\nu/\omega_q)
 \int\! \d^2 \vec k_{q'} \, {B}_{q'/p}( x_{q'}, \vec k_{q'}, \mu,\nu/\omega_{q'})
 \nn\\ 
 &\quad \times
 \int\! \d l_q^+ \int\! \d^2 \vec l_q \, {\tilde {\cal S}}_q(l_q^+, \vec l_q,  \mu,\nu)
 \int\! \d l_{q'}^- \int\! \d^2 \vec l_{q'} \, {\tilde {\cal S}}_{q'}(l_{q'}^-, \vec l_{q'},  \mu,\nu)\nn\\
  &\quad \times
 \int\! \d k\,{\mathcal S}(k,  \mu) \,
\delta^{(2)} \left( {\vec q}_\perp - \vec k_{q} - \vec k_{q'} -\vec l_{q}-\vec l_{q'}\right)
\delta\left( \tau - l_q^+ - l_{q'}^- -k \right)\nn\\
&= \sigma_0\, H(Q, \mu)  \sum_{q,q'} C_{qq'}
\int_0^{\tau_0 Q}\! \d \Tau \int\! \frac{ e^{s \Tau} \d s}{2\pi i}
  \int_0^\infty\! \frac{b\, \d b  }{2\pi} J_0(b\,q_\perp) \, {B}_{q/p}( x_q, b,  \mu,\nu/\omega_q)\,
  {B}_{q'/p}( x_{q'}, b, \mu,\nu/\omega_{q'}) \nn \\
  & \quad \times {\tilde {\cal S}}_q( s, b,  \mu,\nu)\, {\tilde {\cal S}}_{q'}(s, b,  \mu,\nu)\,
 {\mathcal S}( s, \mu) 
\,.
\end{align}
Here, $H(Q,\mu)$ is the hard function, and ${B}_{q/p}(x_q, b,\mu,\nu)$ is the standard TMD beam function, similar to SCET$_{\rm II}$. A new element is ${\tilde {\cal S}}_{a(b)}(s, b,\mu,\nu)$, the double-differential collinear-soft function, which describes collinear-soft emissions sensitive to both 0-jettiness (via $s$) and the transverse momentum (via $b$), respectively. Finally, ${\mathcal S}(s,\mu)$ is the beam thrust soft function, similar to that in SCET$_{\rm I}$. The formula is presented in momentum space and its $b$-space Fourier transform.

The RG equations in $b$-space governing the evolution of these functions are:
\begin{align}
\mu \frac{\d}{\d\mu} {B}_{q/p}( x_q, b, \mu,\nu/\omega_q)&={\gamma}_B^q(\mu,\nu/\omega_q) {B}_{q/p}( x_q, b, \mu,\nu/\omega_q) \,, \nn\\
    \nu \frac{\d}{\d\nu} {B}_{q/p}( x_q, b, \mu,\nu/\omega_q)&=-\frac{1}{2}{\gamma}_\nu^q(b,\mu){B}_{q/p}( x_q, b, \mu,\nu/\omega_q) \,, \nn\\
    \gamma_B^q(\mu,\nu/\omega_q)&=\Gamma_{\rm cusp}^q(\alpha_s) \ln\frac{\nu^2}{\omega_q^2}+\gamma_B^q(\alpha_s)\,,\nn\\
{\gamma}_\nu^q(b,\mu)&=2\left[\int_{\mu^2}^{\mu_b^2}\frac{\d\mu'^2}{\mu'^2}\Gamma_{\rm cusp}^q(\alpha_s)+\gamma_{ r}^q(\alpha_s)\right]\,,\\
    \mu \frac{\d}{\d\mu} {\tilde {\cal S}}_q(s,b, \mu,\nu)
    &={\tilde\gamma}_{\cal S}^q(s,\mu,\nu) {\tilde {\cal S}}_q(s,b, \mu,\nu)  \,, \nn\\
    \nu \frac{\d}{\d\nu} {\tilde {\cal S}}_q(s,b, \mu,\nu) &=\frac{1}{2}{\gamma}_\nu^q(b,\mu) {\tilde {\cal S}}_q(s,b, \mu,\nu)  \,, \nn\\
    {\tilde\gamma}_{\cal S}^q(s,\mu,\nu)&=-2\Gamma_{\rm cusp}^q(\alpha_s) \ln\frac{  \nu}{\mu^2 s e^{\gamma_E}}+{\tilde\gamma}_{\cal S}^q(\alpha_s)\,,\\
    \mu \frac{\d}{\d\mu} {\mathcal S}(s, \mu) &=\gamma_{\mathcal S}^q( s,\mu) {\mathcal S}(s, \mu)  \,, \nn\\
    \gamma_{\mathcal S}^q( s,\mu)&=2\Gamma_{\rm cusp}^q(\alpha_s) \ln\frac{ 1}{\mu^2 s^2 e^{2\gamma_E}}+\gamma_{\mathcal S}^q(\alpha_s)\,,\\
      \mu \frac{\d}{\d\mu} H(Q, \mu) &=\gamma_H^q(Q,\mu) H(Q, \mu)  \,, \nn\\
    \gamma_H^q(Q,\mu)&=2\Gamma_{\rm cusp}^q(\alpha_s) \ln\frac{ Q^2}{\mu^2}+2\gamma_V^q(\alpha_s)\,.
\end{align}
The canonical scales for the functions in the SCET$_{\rm +}$ regime are set as follows:
\begin{align} \label{eq:canonical_scales_scet+}
\mu_H^{\rm +} \sim Q
\,, \qquad
&\mu_B^{\rm +} \sim \mu_b
\,, \qquad
\mu_{\tilde {\cal S}}^{\rm +} \sim \mu_b
\,, \qquad
\mu_{\mathcal S}^{\rm +} \sim \tau_0 Q
\,,\nn\\
&\nu_B^{\rm +} \sim Q
\,, \qquad\ 
\nu_{\tilde {\cal S}}^{\rm +} \sim \frac{\mu_b^2}{\tau_0 Q }
\,.\end{align}

The perturbative Sudakov factor for the SCET$_{\rm +}$ regime is:
\begin{align}
S_{\text P}(b)\!&=\!\frac{C_F}{\pi}\! 
\int_0^{\tau_0 Q}\! \d \Tau \int\! \frac{ e^{s \Tau} \d s}{2\pi i}\left [
 -\int_{\tau_0^2 Q^2}^{\mu_b^2} \!\!
  \frac{\d \mu^2}{\mu^2} 
 \ln \frac{ 1}{\mu^2 s^2 e^{2\gamma_E}}  
 +\! \int_{\mu_b^2}^{Q^2} \!\!
 \frac{\d \mu^2}{\mu^2} \left(
 \ln \frac{ Q^2}{\mu^2}\!-\! \frac{3}{2}\right )   \right ]\!\!  \alpha_s(\mu)  \nn\\
\!&=\!\frac{C_F}{\pi}\! \left [ 
- \int_{\tau_0^2 Q^2}^{\mu_b^2} \!\!
  \frac{\d \mu^2}{\mu^2} 
 \ln \frac{\tau_0^2 Q^2}{\mu^2 }  
 +\! \int_{\mu_b^2}^{Q^2} \!\!
 \frac{\d \mu^2}{\mu^2} \left(
 \ln \frac{ Q^2}{\mu^2}\!-\! \frac{3}{2}\right )   \right ]\!\!  \alpha_s(\mu) \,.   
\end{align}
This factor resums the large logarithms arising from the evolution of the various functions from their respective canonical scales (Eq.~\eqref{eq:canonical_scales_scet+}) to the common scales $\mu_b$.

\subsubsection*{Anomalous Dimensions and Combined Perturbative Sudakov Factor}
The cusp and non-cusp anomalous dimensions are expanded as
\begin{equation} \label{eq:expansion_qnom_dims}
\Gamma^q_\cusp(\as) = \sum_{n = 0}^\infty \Gamma^q_{n} \Bigl( \frac{\as}{4\pi} \Bigr)^{n+1}
\,,\qquad
\gamma^q(\as) = \sum_{n = 0}^\infty \gamma^q_n \Bigl( \frac{\as}{4\pi} \Bigr)^{n+1}
\,.\end{equation}
All the one-loop anomalous dimensions present in the previous section are given by
\begin{align}
  \Gamma^q_{0} &=4C_F\,,\\ 
  \gamma^q_{{\cal B}\,0} &=\gamma^q_{{B}\,0}=6C_F\,,\\ 
  \gamma^q_{{\cal S}\,0} &=\gamma^q_{{S}\,0}={\tilde\gamma}^q_{{\cal S}\,0}=\gamma^q_{{r}\,0}=0\,,\\ 
  \gamma^q_{V\,0} &=-6C_F\,.
\end{align}
To obtain a unified description across the different kinematic regions defined by the scale $\mu_b \sim 1/b$, we combine the results from the three SCET regimes. Specifically:
\begin{itemize}
    \item When $ \mu_b^2 > \tau_0 Q^2$ (small $b$, meaning $q_\perp$ is large relative to $\sqrt{\tau_0}Q$), the DGLAP evolution of PDFs is frozen and instead replaced by scale evolution of beam functions. Hence the SCET$_{\rm I}$ formulation is appropriate, which is consistent with Ref. \cite{Stewart:2009yx}.
    \item When $\tau_0^2 Q^2 < \mu_b^2 < \tau_0 Q^2$ (intermediate $b$), the SCET$_{\rm +}$ formulation is used.
    \item When $\mu_b^2 < \tau_0^2 Q^2$ (large $b$, meaning $q_\perp$ is small relative to $\tau_0 Q$), the SCET$_{\rm II}$ formulation, which corresponds to standard TMD evolution, is applied.
\end{itemize}
We implement the transitions between these regimes using Heaviside $\theta$ functions, resulting in the final combined perturbative Sudakov factor: \begin{align}\label{eq:Sud_SECT}
S_{\text P}(b)\!&=\!\frac{C_F}{\pi}\! \left [
 \int_{\tau_0 Q^2}^{\mu_b^2} \!\!
  \frac{\d \mu^2}{\mu^2} 
 \left(2\ln \frac{\tau_0 Q^2}{\mu^2}-\frac{3}{2}\right) \, \theta(\mu_b^2-\tau_0 Q^2) -
 \int_{\tau_0^2 Q^2}^{\mu_b^2} \!\!
  \frac{\d \mu^2}{\mu^2} 
 \ln \frac{\tau_0^2 Q^2}{\mu^2 } \, \theta(\mu_b^2-\tau_0^2 Q^2) \nonumber \right .\  \\  &
 \quad \!\!\left .\     
 +\! \int_{\mu_b^2}^{Q^2} \!\!
 \frac{\d \mu^2}{\mu^2} \left(
 \ln \frac{ Q^2}{\mu^2}\!-\! \frac{3}{2}\right )   \right ]\!\!  \alpha_s(\mu) \,,    
\end{align}
which is consistent with the Sudakov factor given by Eq.~\eqref{eq:Sud} in the main text. 

\subsection{The supplementary numerical results}
\begin{figure}
\centering
   \hspace{-0.44cm}
 \includegraphics[scale=0.54]{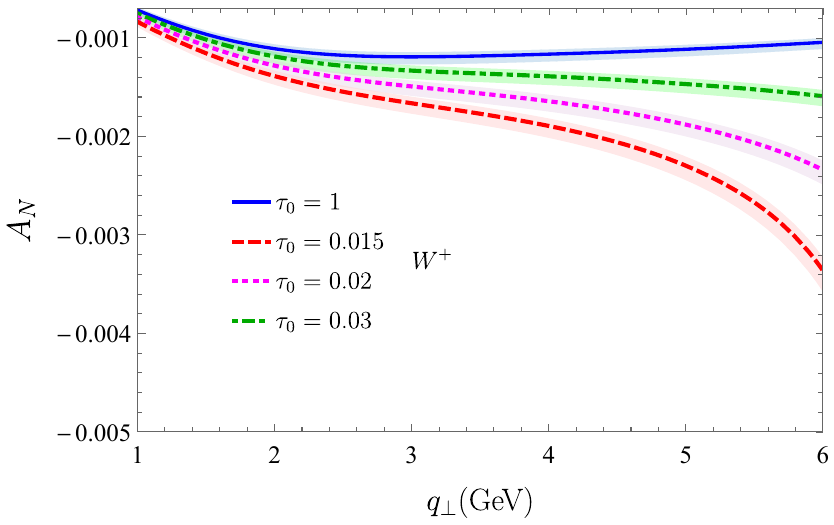}
  \hspace{0cm}
\includegraphics[scale=0.54]{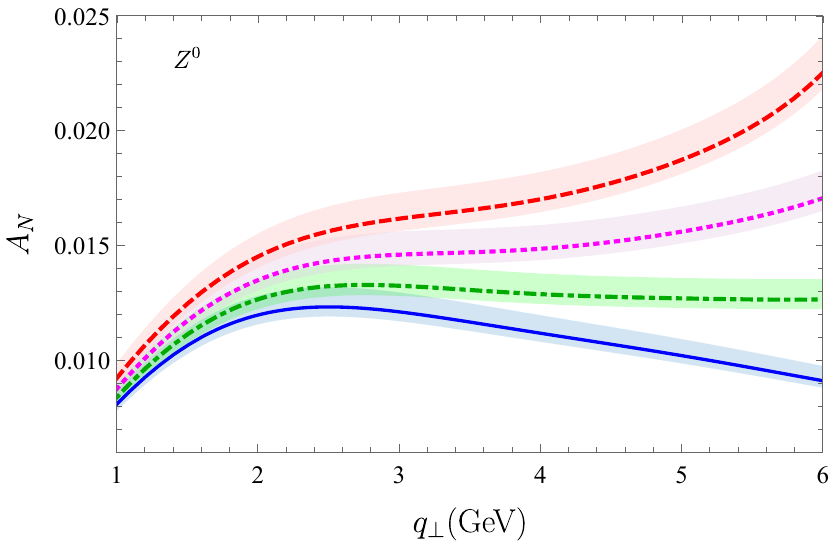}
 \caption{ SSAs for $W^+$ (left) and $Z^0$ (right) production in polarized $pp$ collisions at the RHIC energy $\sqrt{s}=500\,$GeV and the  rapidity $y=0$ are plotted as a function of $q_\perp$ for various values of $\tau_0$. Bands correspond to theoretical uncertainties estimated via scale variation.}
\label{WZ}
\end{figure}

In Fig.~\ref{WZ}, we present numerical predictions for the SSAs in $W^{+}/Z^0$ production as a function of $q_\perp$ at the RHIC energy $\sqrt{s}=500\,$GeV. These asymmetries are computed at mid-rapidity $y=0$ for various values of the 0-jettiness veto parameter $\tau_0$. The $\tau_0=1$ curves correspond to the SSA measured without imposing restrictions on the hadronic final state. As expected, the magnitudes of the SSAs increase as $\tau_0$ decreases. These results demonstrate that implementing a 0-jettiness veto can significantly enhance the sensitivity of SSA measurements to the predicted sign flip of the Sivers function, thereby providing a more robust experimental test of this fundamental QCD prediction.

\begin{figure}
   \centering
 \includegraphics[scale=0.525]{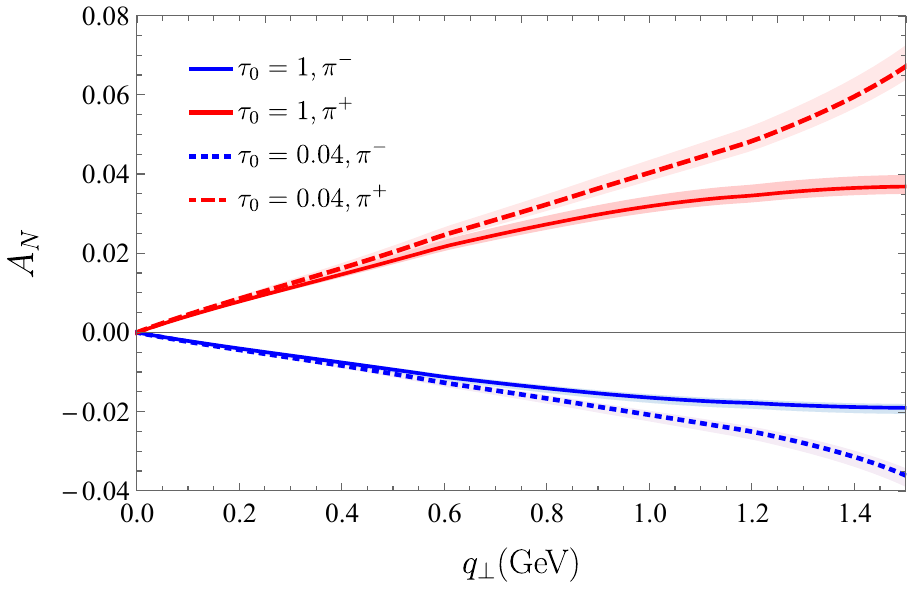}
     \caption{ SSAs for $\pi^+$ and $\pi^-$ production in the SIDIS process at $z_h=0.5$, $x_{B}=0.2$, $Q=25\,$GeV and $\sqrt{s} = 100\,$GeV are plotted as a function of the pion transverse momentum $q_\perp$ for various values of $\tau_0$. The bands show the theoretical uncertainties from scale variation.}
 \label{eic2}
 \end{figure}

In Fig.~\ref{eic2}, we compare the SSAs for $\pi^+$ and $\pi^-$ production in SIDIS at the future EIC, with and without imposing a veto of $\tau_0 = 0.04$. Bands correspond to theoretical uncertainties estimated via scale variation. The results clearly show that the asymmetries are enhanced at moderately large pion transverse momentum when the veto is applied.


\end{document}